\documentclass{aa}
\usepackage{graphicx}
\usepackage[longnamesfirst]{natbib}
\usepackage{txfonts}
\begin{document}
\shortcites{allende1999} \shortcites{bakos2006}\shortcites{baraffe1998} \shortcites{baraffe2003}
\shortcites{blazit1987} \shortcites{bergeron2001} \shortcites{burleigh2002} \shortcites{butler1997}
\shortcites{butler2001} \shortcites{campbell2006}\shortcites{chauvin2006}\shortcites{cushing2005}
\shortcites{cutri2003} \shortcites{danks1994} \shortcites{dasilva2006} \shortcites{desidera2004}
\shortcites{eggenberger2004} \shortcites{eggenberger2006} \shortcites{els2001}
\shortcites{fischer2003} \shortcites{ford2000}\shortcites{fuhrmann2004} \shortcites{halbwachs1986}
\shortcites{hartkopf2000} \shortcites{hartkopf2001} \shortcites{hatzes2003}
\shortcites{holman1999}\shortcites{marcy2005} \shortcites{meyer1998} \shortcites{liebert2005}
\shortcites{kenyon1995} \shortcites{konacki2005} \shortcites{kley2000} \shortcites{lagrange2006}
\shortcites{livio1984} \shortcites{lowrance2002} \shortcites{mayer2005} \shortcites{marzari2000}
\shortcites{mitchell2003} \shortcites{mugrauer2004a} \shortcites{mugrauer2004b}
\shortcites{mugrauer2005a} \shortcites{mugrauer2005b} \shortcites{mugrauer2006}
\shortcites{naef2001b} \shortcites{neuhaeuser2007}\shortcites{patience2002}
\shortcites{perryman1997} \shortcites{pichardo2005} \shortcites{pickles1998}
\shortcites{raghavan2006} \shortcites{randich1999} \shortcites{rocha1998} \shortcites{saffe2005}
\shortcites{santos2004} \shortcites{schmidtkaler1982} \shortcites{soederhjelm1999}
\shortcites{tokovinin2000} \shortcites{weidemann2000} \shortcites{worley1997} \shortcites{wu2003}
\shortcites{zacharias2004} \shortcites{zucker2002b} \shortcites{zucker2003} \shortcites{zucker2004}

\title{The multiplicity of exoplanet host stars \thanks{Based on observations obtained
on La Silla in ESO programs 70.C-0116(A), 71.C-0140(A), 73.C-0103(A), and on Paranal in ESO runs
074.C-0144(B), 074.C-0144(C), 073.C-0370(A), on Mauna Kea in UKIRT program U/02A/16, as well as at
the Munich LMU University Observatory on Mount Wendelstein.}}

\subtitle{Spectroscopic confirmation of the companions GJ\,3021\,B and HD\,27442\,B, one new planet
host triple-star system, and global statistics.}

\author{M. Mugrauer \inst{1} \and R. Neuh\"auser \inst{1} \and T. Mazeh \inst{2}}

\offprints{Markus Mugrauer, markus@astro.uni-jena.de}

\institute{Astrophysikalisches Institut, Universit\"at Jena, Schillerg\"a{\ss}chen 2-3, 07745 Jena, Germany \\
\and Tel Aviv University, Tel Aviv 69978, Israel \\}

\date{Received 22 June 2006 / Accepted 7 February 2007}

\abstract {} {We present new results from our ongoing multiplicity study of exoplanet host stars
and present a list of 29 confirmed planet host multiple-star systems. Furthermore, we discuss the
properties of these stellar systems and compare the properties of exoplanets detected in these
systems with those of planets orbiting single stars.} {We used direct imaging to search for wide
stellar and substellar companions of exoplanet host stars. With infrared and/or optical
spectroscopy, we determined the spectral properties of the newly-found co-moving companions.} {{We
obtained infrared H- and K-band spectra of the co-moving companion GJ\,3021\,B. The infrared
spectra and the apparent H-band photometry of the companion is consistent with an M3 -- M5 dwarf at
the distance of the exoplanet host star.

HD\,40979\,AB is a wide planet host stellar system, with a separation of $\sim$\,6400\,AU. The
companion to the exoplanet host star turned out to be a close stellar pair with a projected
separation of $\sim$130\,AU, hence, this system is a new member of those rare planet host
triple-star systems of which only three other systems are presently known.

HD\,27442\,AB is a wide binary system listed in the Washington Double Star Catalogue, whose common
proper motion was recently confirmed. This system is composed of the subgiant HD\,27442\,A hosting
the exoplanet, and its faint companion HD\,27442\,B. The visible and infrared J-, H-, and
K$_{S}$-band photometry of HD\,27442\,B at the distance of the primary star shows that the
companion is probably a white dwarf. Our multi-epochs SofI imaging observations confirm this result
and even refine the suggested physical characteristics of HD\,27442\,B. This companion should be a
relatively young, hot white dwarf with an effective temperature of $\sim$14400\,K, and cooling age
of $\sim$220\,Myr. Finally, we could unambiguously confirm the white dwarf nature of HD\,27442\,B
with follow-up optical and infrared spectroscopy. The spectra of the companion show Hydrogen
absorption features of the Balmer, Paschen, and Bracket series. With its subgiant primary and the
white-dwarf companion, the HD\,27442\,AB system is the most evolved planet host stellar system
known today.

The mass-period and eccentricity-period correlation of planets around single stars and those
residing in multiple-star systems seem different for the short-period planets. The distribution
functions of planet orbital elements (P, e) are identical, while the mass-distribution ($msin(i)$)
exhibits one difference. While both planet populations exhibit a peak in their mass-distribution at
about 1\,$M_{Jup}$, the frequency of more massive planets continually decreases in single-star
systems, whereas the mass-distribution of planets residing in multiple-star systems exhibits a
further peak at about 4\,$M_{Jup}$. This indicates that the mass-distributions of the two planet
populations might differ in the intermediate mass-range between 2 and 6\,$M_{Jup}$.}}{}


\maketitle

\section{Introduction}

Planets are believed to form in a gas and dust disk by either gravitational disk instability or
accretion. If the planet-bearing disk surrounds a star that is located in a multiple-star system,
the stellar companion(s) not only truncate(s) the extent of the disk \citep[see
e.g.][]{pichardo2005} but also directly alter(s) the planet formation processes. Therefore, the
characteristics of planets in multiple-star systems might be different from those of planets
orbiting single stars.

For example, \citet{mayer2005} study the influence of the perturbations of the stellar companion on
the gravitational instability scenario of the planet-bearing disk by comparing disks with different
properties around single stars to disks in a binary system, with a perturbing stellar companion.
The tidal interaction of the stellar disturber induces spiral-like arms in the disks, i.e. regions
of increased gas density and temperature.

In particular, intermediate-mass disks ($\sim$\,50\,$M_{Jup}$) that are gravitationally stable in
isolation start to fragment in multiple-star systems, and over-dense gas clumps form along the
induced spiral arms, clumps that finally lead to planet formation via disk instability. On the
other hand, in more massive disks the induced compressional heating along the dense spiral arms is
so strong that the disk remains gravitationally stable, although it would fragment in a single-star
system. Hence, stellar multiplicity may either trigger or hamper planet formation via gravitational
disk instability, depending on the characteristics of the stellar system, as well as on the disk
properties (mass, cooling times).

In the accretion scenario, dust particles accumulate into planetesimals that finally accrete into
rocky earth-like planets, while the cores of gas giant planets are formed later on. The accretion
process is very sensitive to the relative velocities of the interacting particles. In order that
accretion can occur, the relative velocities must not be too high, which would favor destruction of
colliding particles. On one hand, a stellar companion induces high particle eccentricities and
therefore the relative velocities of colliding objects are substantially increased. On the other
hand, there are also dissipative mechanisms (most important here is the gas drag force) that again
reduce the induced, high relative velocities. Whether accretion to larger bodies occurs or
fragmentation of colliding objects into dust is favored depends on which of these two processes
dominates \citep{marzari2000}. Therefore, stellar multiplicity also influences the efficiency of
planet formation in the accretion scenario.

Once formed, giant gas planets that remain embedded in the disk continue to accrete material and
might also migrate within the disk. According to \citet{kley2000}, stellar companions alter both
processes by varying the density structure of the protoplanetary disk. The accretion is enhanced
and the migration time reduced, which also damps the orbital eccentricities of the migrating
planets.

Finally, after the disappearance of the planet bearing disk, the orbital elements of planets in
binary systems can be altered over a long span of time by the Kozai effect, which oscillates the
planet eccentricity and can also lead to an orbital migration if the planetary orbital plane is
sufficiently inclined to the orbital plane of the stellar system; see. e.g. \cite{ford2000} and
\cite{wu2003}. Thus, the orbital characteristics of planets residing in multiple-star systems and
those orbiting around single stars could show differences that might shed some light on their
formation and evolution.

During the decade of extrasolar intensive study, many planets in multiple-star systems were
discovered. Already during the first two years of the detection of the first low-mass exoplanet
51\,Peg\,b \citep{mayor1995}, \citet{butler1997} found three exoplanets residing in double-star
systems --- 55\,Cnc, $\tau$\,Boo, and $\upsilon$\,And. While 55\,Cnc and $\tau$\,Boo were both
already known as binary systems, $\upsilon$\,And was considered as a single star at that time.
Later on, \citet{lowrance2002} found a co-moving wide companion of $\upsilon$\,And, turning it into
a planet host binary system. Both $\upsilon$\,And and 55\,Cnc are wide binaries composed of the
bright exoplanet host star and a faint low-mass M-dwarf companion, separated from the planet host
star by several hundred AU. The majority of all planet host binaries known today are widely
separated pairs like 55\,Cnc and $\upsilon$\,And. In contrast, $\tau$\,Boo belongs to the group of
those few relatively close planet host binary systems with separations smaller than 100\,AU. Only
six of those close stellar systems are presently known (see Table\,\ref{table8}).

The apparent lack of close binaries hosting planets is probably a selection effect due to the
difficulty of detecting planets in close binary systems using the radial-velocity technique (see
Sect.\,4). If the two components of a binary system cannot be spatially separated by the
spectrograph, only the sum of both stellar spectra can be detected (double-lined spectroscopic
binary). Even in the case of a high flux ratio between primary and secondary
(F$_{Prim}$/F$_{Sec}$\,$\lesssim$\,100), the primary spectrum is affected by the parasitic light of
the secondary, inducing changes in the line profiles of the primary spectrum. To avoid these
perturbations in the planet search, close stellar systems are excluded from the large planet search
campaigns studying several thousand F to K stars. Only wide binaries, in which one can take spectra
of each component separately, or close systems composed of a bright primary and a much fainter
secondary star are targets of these programs.

Nevertheless, planet searches in close binary systems is possible. One can model the spectrum of
the system with the sum spectrum of the two components with different radial velocities, as is done
by TODCOR (Zucker and Mazeh 1994) or TODMOR \citep[][2004]{zucker2003}. One can use this technique
to finally derive the radial velocities of both components, allowing a planet search around both
binary components. Today radial-velocity projects are on the way toward searching for planets in
wide \citep[][; Toyota et al. 2005\nocite{toyota2005}]{desidera2004} as well as in close
(double-lined spectroscopic) binary systems \citep{konacki2005a}. A first promising result was
reported by \citet{konacki2005b}, who found a planet orbiting the primary component of the close
triple-star system HD\,188753\,A/B+C.

Since the detection of the first exoplanets in binaries \citep{butler1997}, more and more exoplanet
host stars, once considered as single stars, turned out to be components of multiple-star systems,
detected in imaging programs that search for visual companions of exoplanet host stars. In order to
find both close and wide companions, these imaging campaigns are carried out with high-contrast,
small-field adaptive optics and with normal seeing-limited imaging (large field of view).
\citet{patience2002} detected a close companion to HD\,114762. Furthermore, \citet{els2001} report
on a faint companion only 2\,arcsec separated from Gl\,86, that eventually turned out to be a white
dwarf (Jahrei\ss\,\,2001 \nocite{jahreiss2001}; Mugrauer \& Neuh\"auser 2005
\nocite{mugrauer2005a}; Lagrange et al. 2006\nocite{lagrange2006}). Finally, Mugrauer et al.
(2004a, 2004b, 2005, 2006) \nocite{mugrauer2004a} \nocite{mugrauer2004b} \nocite{mugrauer2005b}
\nocite{mugrauer2006} present 7 new wide companions of exoplanet host stars, namely HD\,89744,
HD\,75289, HD\,16141, HD\,114729, HD\,196050, HD\,213240, and HD\,46375.

In this paper we publish additional results of our multiplicity study. We present our imaging and
photometric data for the companions GJ\,3021\,B, HD\,27442\,B \citep[see also][Raghavan et al.
2006\nocite{raghavan2006}]{chauvin2006} and report a new co-moving companion to the planet host
star HD\,40979 in Sect.\,2. In Sect.\,3 we summarize the results of the infrared and optical
follow-up spectroscopy of GJ\,3021\,B and HD\,27442\,B. In Sect.\,4 we compile an updated list of
confirmed planet host multiple-star systems and discuss their properties. Finally, in Sect.\,5, we
compare the masses and orbital parameters of planets residing in multiple-star systems with those
of planets orbiting single stars.

\section{Astrometry and photometry}

GJ\,3021 and HD\,27442 are targets of our southern search program for the visual companions of
exoplanet host stars. Observations were carried out with the infrared camera SofI \footnote{Son of
ISAAC}, a 1024\,$\times$\,1024 HgCdTe-detector installed at the Nasmyth focus of the 3.58\,m
NTT\footnote{New Technology Telescope} telescope. To be sensitive to faint substellar companions,
all targets were observed in the near infrared H-band at 1.6\,$\mu$m, where companions with masses
as low as a few tens of Jupiter masses come into the range of a 4\,m class telescope. Most of the
exoplanet host stars are located within a distance of about 60\,pc around the sun, thus all are
relatively bright sources in the night sky. In order to avoid strong saturation effects, we used
the smallest available pixel scale and shortest available detector integration times. We observed
all targets in the SofI small-field mode with its pixel scale of 144\,mas per pixel, yielding a
field of view of 147\,arcsec\,$\times$\,147\,arcsec. The spacial resolution samples the typical
seeing conditions well, which are better than 1\,arcsec at the La Silla ESO Observatory. Therefore,
the proper motions of detected objects can be determined after one year with a precision of 20 to
30\,mas over the whole SofI field of view. Besides the detection of wide companions, the small
pixel scale also allows the detection of close companions (see e.g. GJ\,3021\,B), which are
separated from the bright star by only a few arcsec. Only very close companions with a separation
of less than about 2\,arcsec remain hidden in the glare of the primary star. Those can be revealed
only by high-contrast AO imaging \citep[see e.g.][]{mugrauer2005a}.

We took 10 images per target, each of which is an average of 50$\times$1.2\,s integrations, i.e.
10\,min of total integration time per target. In order to cancel out the high infrared background
we applied the standard jitter technique; i.e. the telescope is moved by a few arcsec to a new
position before the next image is taken. Hence, each object is located at a different position on
the detector at the different images, a fact that allows the determination of the infrared
background at each pixel. The background estimation and subtraction, as well as the flat-fielding
of all images, is achieved with the ESO package \textsl{ECLIPSE}\footnote{ESO C Library for an
Image Processing Software Environment} \citep{devillard2001}, which finally also combines all
images, using a shift+add procedure.

The astrometry of all SofI images is calibrated with the 2MASS\footnote{2 Micron All Sky Survey}
point source catalogue \citep{cutri2003}, which contains accurate positions of objects brighter
than 15.2\,mag in H ($S/N$\,$>$\,5). The derived pixel scale $PS$ and detector orientation $PA$ of
all observing runs, whose data are presented in this paper, are summarized in Table\,\ref{table1}.
The SofI detector is tilted by $PA$ from north to west.

\begin{table}[ht]
\caption{The astrometrical calibration of SofI, UFTI (IR), and MONICA (Optical) in the individual
observing runs.}
\begin{center}
\begin{tabular}{c|c|c|c}
\hline\hline
instrument & epoch & $PS$ [arcsec/pixel] & $PA$ [$^{\circ}$]\\
\hline
SofI$_{small}$ & 12/02 & 0.14366$\pm$0.00016 & 90.069$\pm$0.041\\
SofI$_{small}$ & 06/03 & 0.14365$\pm$0.00013 & 90.069$\pm$0.032\\
SofI$_{small}$ & 07/04 & 0.14356$\pm$0.00011 & 90.047$\pm$0.024\\
\hline
UFTI & 11/02 & 0.09098$\pm$0.00043 &  \,\,\,0.770$\pm$0.087\\
UFTI & 10/03 & 0.09104$\pm$0.00030 &  \,\,\,0.711$\pm$0.083\\
\hline
MONICA & 11/04 & 0.49051$\pm$0.00015 & 0.003$\pm$0.035\\
MONICA & 07/05 & 0.49042$\pm$0.00029 & $-$0.013$\pm$0.063\\
\hline\hline
\end{tabular}
\label{table1}
\end{center}
\end{table}

\subsection{GJ\,3021}

GJ\,3021 (HD\,1237) is a nearby G6V dwarf \citep[$\sim$18\,pc Hipparcos]{perryman1997}, located in
the southern constellation Hydrus. \citet{naef2001b} detected a massive exoplanet
($msin(i)$\,=\,3.37\,$M_{Jup}$) with a period of 134 days and an eccentric (e\,=\,0.511) orbit with
a semi-major axis of 0.49\,AU. According to \citet{naef2001b}, the star is a photometric variable
and exhibits a high level of chromospherical activity. Its spacial velocity ($U$\,=\,33.7\,km/s,
$V$\,=\,17.4\,km/s, $W$\,=\,2.8\,km/s) metallicity $[Fe/H]$\,=\,0.1$\pm$0.08, chromospheric
activity level $log({R^{'}_{HK}})$\,=\,$-$4.27, as well as its lithium abundance
$log(N(Li))$\,=\,2.11$\pm$0.08, are all typical of members of the Hyades super cluster. The
chromosperical age ranges from 0.02 to 0.8\,Gyr, which is consistent with an age of 0.5\,Gyr, as
derived by \cite{rocha1998} and with the age estimate of 0.15 to 0.25\,Gyr from \citet{saffe2005}.

\begin{figure} [h]
\resizebox{\hsize}{!}{\includegraphics{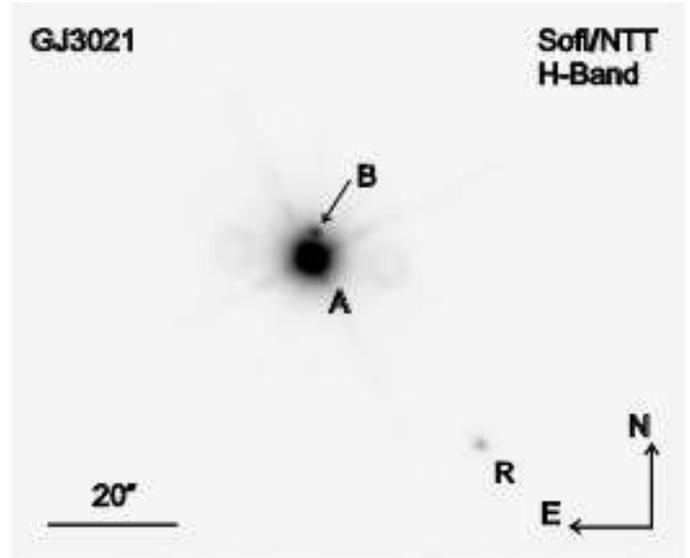}}\caption{The SofI small-field image of the
planet host star GJ\,3021 (central bright star A with diffraction spikes and reflections), taken in
December 2002 in the H-band, with a total integration time of 10 min. Only two companion candidates
are detected with $S/N$\,$\ge$\,10 close to the exoplanet host star. GJ\,3021\,B is co-moving to
the exoplanet host star. The star R, whose proper motion is listed in the UCAC2 catalogue is used
as the astrometric reference.}\label{fig1}
\end{figure}

We obtained a first image of GJ\,3021, shown in Fig.\,\ref{fig1}, in December 2002. Follow-up,
second-epoch observations were carried out in June 2003. Only two companion candidates were
detected with $S/N$\,$\ge$\,10 in the SofI images. The closer one (B) is located only
$\sim$4\,arcsec north of GJ\,3021, and the other candidate (R) is detected $\sim$39\,arcsec
southwest of the exoplanet host star. Both objects could be either real companions of the exoplanet
host star or just background stars randomly located close to but far behind the exoplanet host
star. Because of its proximity to the sun, GJ\,3021 exhibits a high proper motion
($\mu_{\alpha}cos(\delta)$\,=\,433.88$\pm$0.55\,mas/yr,
$\mu_{\delta}$\,=\,$-$57.91$\pm$0.45\,mas/yr), derived from precise measurements of the European
astrometry satellite Hipparcos \citep{perryman1997}, which also provided the parallax of the star
($\pi$\,=\,56.76$\pm$0.53\,mas). Real companions of the exoplanet host star must share the proper
motion of their primary stars, as their orbital motion is much smaller than the proper motion. Such
co-moving companions can therefore be distinguished from non-moving or slowly moving background
stars by comparing two images taken with a sufficiently long time difference.

The more distant star, R, is listed in the UCAC2 catalogue \citep{zacharias2004} with a proper
motion $\mu_{\alpha}cos(\delta)$\,=\,1.8$\pm$5.9\,mas/yr and
$\mu_{\delta}$\,=\,$-$24.9$\pm$6.1\,mas/yr, which is significantly different from the proper motion
of the exoplanet host star. Hence, we can clearly rule out the companionship of this candidate,
which is only a slow-moving object, most probably located in the background.

Recently, \cite{chauvin2006} have also shown that the closer companion candidate (HD\,1237\,B)
shares the proper motion of the planet host star. By comparing the two SofI images, we can confirm
this astrometric detection. The measured proper motion of the companion between the two SofI
observing epochs is $PM_{Ra}$\,=\,368$\pm$68\,mas, $PM_{Dec}$\,=\,$-$48$\pm$68\,mas, which is
consistent with the expected proper motion of the exoplanet host star
($PM_{Ra}$\,=\,317$\pm$1\,mas, $PM_{Dec}$\,=\,$-$6$\pm$1\,mas) for the given timespan. The
separation and position angle of GJ\,3021\,B relative to its primary at both SofI observing epochs
are summarized in Tab.\,\ref{table3}.

\begin{table}[htb]
\caption{The separations and position angles of GJ\,3021\,B and HD\,27442\,B relative to their
primaries for all SofI observing epochs.}
\begin{center}
\begin{tabular}{c|c|c|c}
\hline\hline
GJ\,3021\,B & epoch      & sep[arcsec]       & PA[$^{\circ}$]\\
\hline
            & NTT 12/02  &  3.892$\pm$0.076  & 355.480$\pm$0.858\\
            & NTT 06/03  &  3.810$\pm$0.075  & 354.570$\pm$0.902\\
\hline\hline
HD\,27442\,B & epoch      & sep[arcsec]       & PA[$^{\circ}$]\\
\hline
             & NTT 12/02  &  12.933$\pm$0.083 & 36.400$\pm$0.364\\
             & NTT 07/04  &  12.940$\pm$0.079 & 36.980$\pm$0.347\\
\hline\hline
\end{tabular}
\end{center}\label{table3}
\end{table}

By measuring the noise level in the individual SofI images of both observing runs, we derived the
detection limits of the SofI observations. Figure\,2 shows the achieved detection limit
(S/N\,=\,10) for a range of separations to the planet host star for a system age of 0.5\,Gyr. Due
to the low declination of GJ\,3021, our SofI observations were always carried out at high airmasses
(AM$\ge$1.5) with an average FWHM of only $\sim$1.5\,arcsec. A limiting magnitude of 16.7\,mag is
reached in the background-limited region, i.e. at angular separations around GJ\,3021\,A larger
than 15\,arcsec. If we assume a system age of 0.5\,Gyr, the achieved limiting magnitude enables the
detection of brown-dwarf companions with a mass m\,$\ge$\,19\,$M_{Jup}$, derived with the
\citet{baraffe2003} models and their magnitude-mass relation. This mass-limit increases for higher
system ages, e.g. 56\,$M_{Jup}$ at 5\,Gyr.

\begin{figure} [htb]
\resizebox{\hsize}{!}{\includegraphics{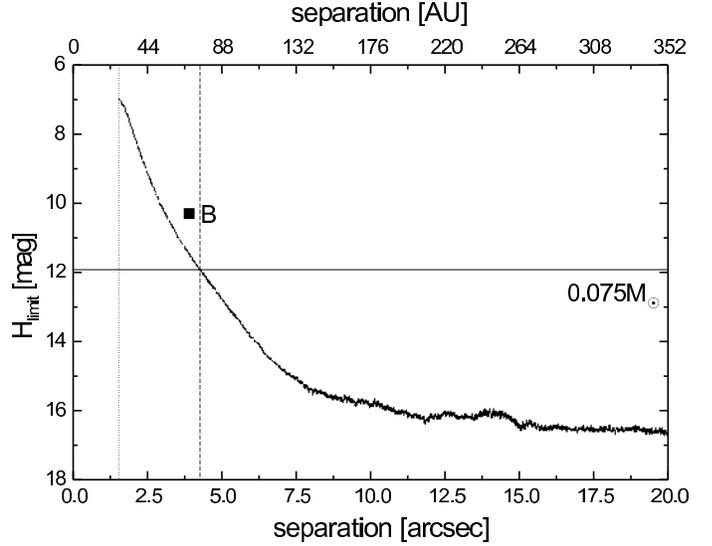}}\caption{The detection limit ($S/N$\,=\,10)
of the SofI H-band imaging of GJ\,3021 for a range of separations given in arcsec at the bottom and
as projected separation in AU at the top. Saturation occurs at $\sim$1.5\,arcsec (dotted line),
i.e. companions with a projected separation closer than 26\,AU are not detectable. At a system age
of 0.5\,Gyr, all stellar companions (m\,$\ge$\,75\,$M_{Jup}$) are detectable beyond the distance
illustrated by the dashed line ($\sim$74\,AU).}
\end{figure}\label{fig2}

Objects with angular separations of up to 63\,arcsec ($\sim$1100\,AU) around the exoplanet host
star are imaged twice in both SofI observing runs. The SofI field of view is about 4.5 times larger
than the field of view investigated by \cite{chauvin2006} using NACO and its S27 camera. Wide
companion candidates, which lie outside the small AO field of view (see e.g. object R), can be
detected only with SofI. In contrast, faint close companions that need higher contrast can only be
found with adaptive optics, like NACO. Hence, both the seeing-limited and the diffraction-limited
observations yield complementary information about the multiplicity of the planet host star for
different ranges of angular separation. In the case of GJ\,3021\,A, our SofI images can rule out
any other wide co-moving companion, expect GJ\,3021\,B. Combining our SofI imaging data with the
NACO observations obtained by \cite{chauvin2006}, additional stellar companions
(m$\ge$75\,$M_{Jup}$) can be ruled out around GJ\,3021\,A with projected separations from
$\sim$20\,AU up to $\sim$1110\,AU.

We measured the companion photometry in both SofI images and obtained an apparent H-band magnitude
H\,=\,10.298$\pm$0.054\,mag, consistent with the NACO photometry presented by \cite{chauvin2006}.
With the given distance of the exoplanet host star, we obtained an absolute H-band magnitude of the
companion M$_{H}$=9.068$\pm$0.058\,mag. According to the magnitude-mass relation of the
\citet{baraffe1998} evolutionary models and the color-spectral type conversion from
\citet{kenyon1995}, the companion photometry is consistent with an M4-M5 dwarf with a mass of
0.125$\pm$0.003\,$M_{\odot}$ at an age of 0.5\,Gyr. This estimate, derived only by photometry,
still has to be confirmed by spectroscopy (see Sect.\,3).

\subsection{HD\,27442}

HD\,27442 ($\epsilon$\,Ret) is $\sim$10\,Gyr old K2 subgiant \citep{randich1999} located at a
distance of $\sim$18\,pc ($\pi=54.84\pm0.50$\,mas) from the sun, derived by Hipparcos
\citep{perryman1997}. According to \citet{santos2004} HD\,27442 exhibits a surface gravity of
$log(g)$\,=\,3.55$\pm$0.32\,cms$^{-2}$. The age estimate of the star, as well as its surface
gravity, is fully consistent with the subgiant classification. \citet{butler2001} detected an
exoplanet ($msin(i)$\,=\,1.35\,$M_{Jup}$) in an almost circular orbit (e\,=\,0.058) around
HD\,27442 with an orbital period of 415 days and a semi-major axis of 1.16\,AU.

\begin{figure} [h]
\resizebox{\hsize}{!}{\includegraphics{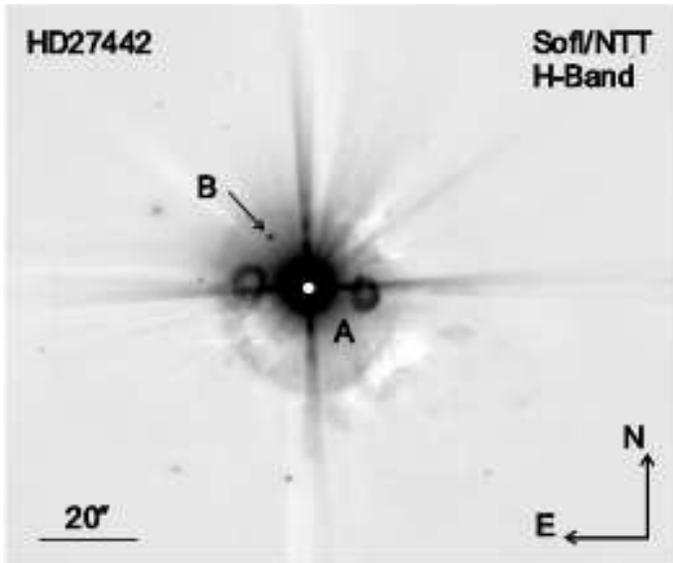}}\caption{The first-epoch SofI small-field image
of the exoplanet host star HD\,27442 observed in December 2002. The total integration time of this
H-band image is 10\,min. Several faint companion candidates were detected ($S/N>10$). The co-moving
companion HD\,27442\,B is visible about 13\,arcsec northeast of the bright exoplanet host
star.}\label{fig3}
\end{figure}

HD\,27442 is listed in the WDS\footnote{Washington Double Star} catalogue \citep{worley1997} as a
binary system with a V\,=\,12.5\,mag companion, based on two astrometric measures in 1930 and one
follow-up observation in 1964. In 1930 the companion was located at a separation of
13.7$\pm$0.1\,arcsec and at a position angle of 35$\pm$1\,$^{\circ}$. Recently, \cite{chauvin2006}
and \cite{raghavan2006} also reported on their observations of the HD\,27442 system. Both groups
could clearly detected the companion HD\,27442\,B close to its primary HD\,27442\,A and confirmed
that HD\,27442\,A and B form a common proper-motion pair. We observed HD\,27442 with SofI in
December 2002 and July 2004. Our first-epoch H-band image is shown in Fig.\,\ref{fig3}. After
10\,min of total integration, several faint companion candidates were detected close to the bright
exoplanet host star. The co-moving companion HD\,27442\,B is shown in Fig.\,\ref{fig3}, and the
SofI astrometry of HD\,27442\,B relative to the planet host star is summarized in
Table\,\ref{table3}.

\begin{figure} [h]
\resizebox{\hsize}{!}{\includegraphics{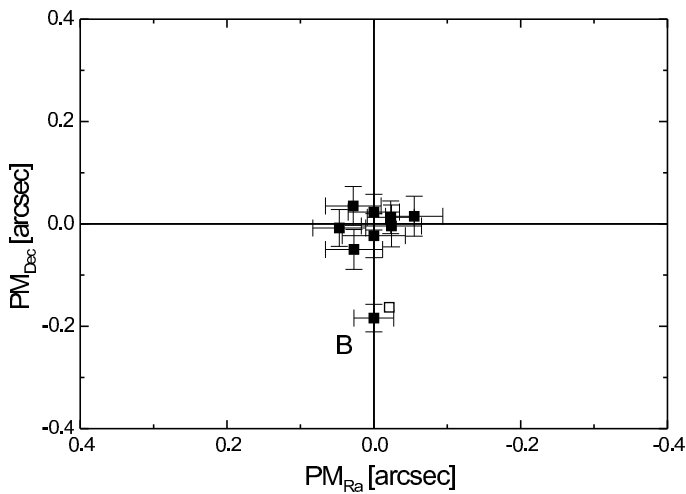}}\caption{The derived proper motion of all
companion candidates detected in our NTT images (S/N$>$10) from epochs 12/02 and 07/04. The
expected proper motion of the primary is derived from Hipparcos data and is shown as a small white
square in the plot.}\label{fig4}
\end{figure}

In our SofI H-band images, several wide companion candidates were detected that are separated from
the bright planet host star by up to 64\,arcsec ($\sim$1200\,AU). By comparing the first-epoch with
the second-epoch SofI image, the proper motions of all these companion candidates can be determined
for the given epoch difference and are shown in Fig.\,\ref{fig4}. HD\,27442\,B clearly shares the
proper motion of HD\,27442\,A, confirming the astrometric measurements of \cite{chauvin2006} and
\cite{raghavan2006}. Expect for HD\,27442\,B, no additional co-moving companions could be
identified around the exoplanet host star within the SofI field of view.

\begin{figure} [h]
\resizebox{\hsize}{!}{\includegraphics{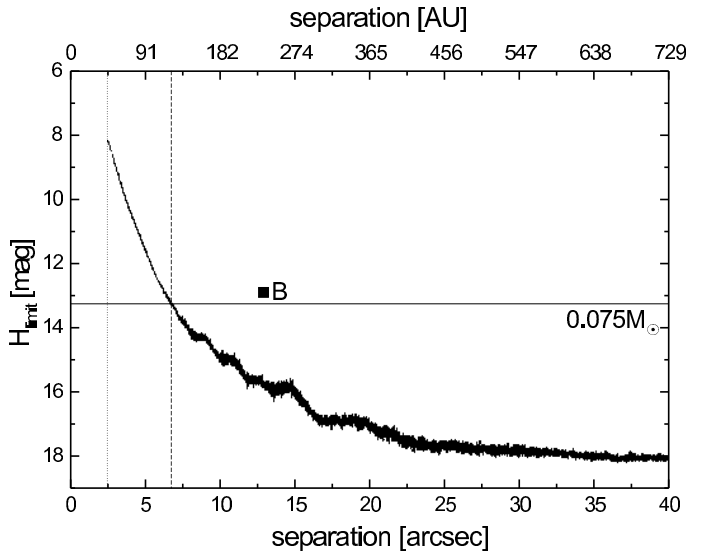}}\caption{The detection limit ($S/N$\,=\,10)
of our SofI H-band images of HD\,27442 for a range of angular separations given in arcsec at the
bottom and as projected separations in AU at the top. At $\sim$2.5\,arcsec saturation occurs (doted
line), i.e. companions with a projected separation closer than 46\,AU are not detectable. At a
system age of 10\,Gyr all stellar companions ($m$\,$\ge$\,0.075\,$M_{\sun}$) are detectable beyond
6.8\,arcsec ($\sim$125\,AU), as illustrated by a dashed line.}\label{fig6}
\end{figure}

The SofI detection limit versus distance to HD\,27442\,A is illustrated in Fig.\,\ref{fig6}. In
both SofI images, the seeing is $\sim$0.8\,arcsec and a limiting magnitude ($S/N$\,=\,10) of
H\,=\,18.0\,mag is reached beyond $\sim$30\,arcsec (547\,AU) in the background-limited region.
According to \citet{baraffe2003}, the limiting magnitude allows the detection of substellar
companions with masses down to 0.055\,$M_{\sun}$ at the approximated age of the exoplanet host star
of 10\,Gyr. With the high-contrast AO observations obtained by \cite{chauvin2006} and our SofI
wide-field imaging data of the HD\,27442\,AB system, additional stellar companions of the exoplanet
host star, with projected separations between $\sim$30\,AU and 1200\,AU, can be ruled out.

\cite{chauvin2006} combined their J- and K$_{S}$-band NACO photometry of HD\,27442\,B with its
visible magnitude listed in the WDS catalogue and concluded that the companion photometry is
inconsistent with a main sequence star or a brown dwarf at the distance of the exoplanet host star.
Instead, the photometry of the co-moving companion is consistent with what is predicted by the
evolutionary model of \cite{bergeron2001} for white dwarfs with hydrogen- or helium-rich
atmospheres, with a mass ranging between 0.3 and 1.2\,$M_\odot$ and an effective temperature
ranging between 9000 and 17000\,K.

\begin{figure} [h]
\resizebox{\hsize}{!}{\includegraphics{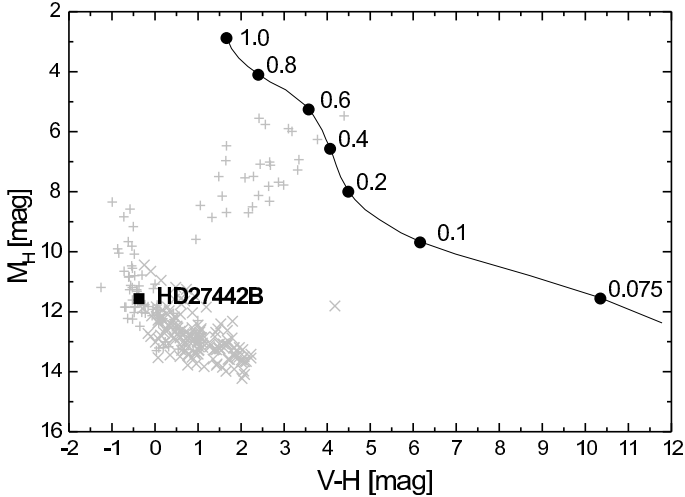}}\caption{The M$_{H}$$-$(V$-$H) color-magnitude
diagram for main-sequence stars, white dwarfs, and HD\,27422\,B. The 8\,Gyr isochrone from the
\citet{baraffe1998} models is shown as a solid line, plotted for stellar masses from 0.072 to
1\,$M_{\sun}$. The masses of objects on the isochrone are indicated with black circles. For
comparison we show the location of white dwarfs from \citet{bergeron2001} (X symbols) and from
\citet{liebert2005} (+ symbols).}\label{fig5}
\end{figure}

We can confirm the white dwarf hypothesis by comparing the visible WDS and SofI H-band photometry
of HD\,27442\,B with known white dwarfs. The average H-band photometry of HD\,27442\,B in the two
SofI images is H\,=\,12.871$\pm$0.085\,mag, which is consistent with a
M$_{H}$\,=\,11.566$\pm$0.087\,mag object at the distance of the exoplanet host star. With
V\,=\,12.5$\pm$0.1\,mag from the WDS catalogue, this yields the color of the companion
V$-$H\,=\,$-$0.37$\pm$0.13\,mag. Figure\,\ref{fig5} shows HD\,27442\,B in a color-magnitude
(M$_{H}$-$V$-$H)$ diagram, together with the known white dwarfs from the Palomar Green Survey
\citep{liebert2005}, and old white dwarfs from \citet{bergeron2001}. Most of the white dwarfs are
well-separated from the isochrone of low-mass stellar objects. They are faint and much bluer than
low-mass stellar objects with a similar brightness. Only the youngest (age$\le$0.29\,Gyr), hottest
(T$\le$70000\,K), and therefore almost brightest white dwarfs in the comparison sample extend to
the upper-right corner of the diagram reaching the stellar isochrone. These relatively young white
dwarfs are brighter and redder than their older analogues.

The photometry of HD\,27442\,B is consistent with a white dwarf companion at the distance of the
exoplanet host star. We find several white dwarfs in the Palomar Green Survey with V- and H-band
magnitudes that are comparable with those of HD\,27442\,B. The properties of these white dwarfs are
summarized in Table\,\ref{table4}.

\begin{table}[ht]
\caption{Properties of comparison white dwarfs from the Palomar Green Survey whose absolute V- and
H-band photometry is comparable to HD\,27442\,B.}
\begin{center}
\begin{tabular}{l|c}
\hline\hline
$T_{eff}$ & 14400$\pm$3100\,K\\
$log(L[L_{\sun}])$ & $-$2.09$\pm$0.22\,\,\,\,\,\,\\
$log(g[cms^{-2}])$ & 7.95\,$\pm$0.17\,\,\,\,\\
$log(\tau[yr])$ & \,8.35$\pm$0.14\,\,\,\,\\
mass & \,\,\,\,\,\,\,0.58$\pm$0.11\,$M_{\sun}$\\
\hline\hline
\end{tabular}
\end{center}\label{table4}
\end{table}

Finally, the white dwarf nature of HD\,27442\,B, derived so far only from colors and absolute
magnitudes, has to be confirmed with spectroscopy. Therefore, we obtained follow-up spectra of the
companion, which are presented in Sec.\,3.

\subsection{HD\,40979}

HD\,40979 is located in the northern constellation Auriga. This star is a nearby F8 dwarf at a
distance of 33\,pc, as derived by Hipparcos \citep{perryman1997}. According to \citet{fischer2003}
this star shows a strong lithium absorption feature $logN(Li)$\,=\,2.79 but exhibits only a
moderate chromospheric activity $logR'_{HK}$\,=\,$-$4.63. They conclude that the lithium abundance
of HD\,40979, its color B$-$V\,=\,0.537$\pm$0.007\,mag \citep[data from Hipparcos
see][]{perryman1997}, as well as its Ca H and K emission, are all consistent with a 1.1 to
1.2\,$M_{\sun}$ dwarf at an age of about 1.5\,Gyr. Furthermore, they report a periodical modulation
of the stellar radial velocity with a period of 263\,days. They suggest that this variation is
induced by an exoplanet ($msin(i)$\,=\,3.28\,$M_{Jup}$) that orbits the star on an eccentric
(e\,=\,0.25) orbit with a semi-major axis of 0.83\,AU.

HD\,40979 is a target of our imaging search campaign for visual companions of northern exoplanet
host stars, which is being carried out with the 3.8\,m United Kingdom Infrared Telescope (UKIRT),
located at Mauna Kea (Hawaii). We took H-band images of all northern targets with the infrared
camera UFTI\footnote{UKIRT Fast Track Imager}, which is equipped with a 1024\,$\times$\,1024 HgCdTe
infrared detector, providing a pixel scale of $\sim$91\,mas per pixel and a
93\,arcsec\,$\times$\,93\,arcsec field of view. In order to reduce saturation effects we always
used the shortest possible detector integration time (4\,s) and add up 6 integrations per jitter
position. We always chose 24 jitter positions, yielding a total integration time of 9.6\,min per
target. The data reduction was carried out again with ESO \textsl{ECLIPSE} \citep{devillard2001}.

\begin{figure} [h]
\resizebox{\hsize}{!}{\includegraphics{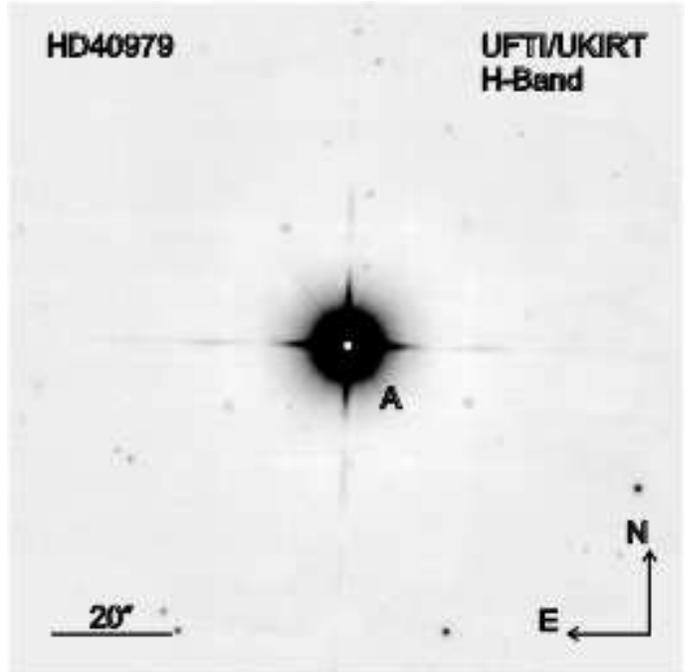}}\caption{Our first-epoch UFTI image of the
exoplanet host star HD\,40979, observed in November 2002 in the H-band. Within the UFTI field of
view, several faint companion candidates ($S/N$\,$>$\,10) were detected around the exoplanet host
star. The proper motion of those candidates are shown in Fig.\ref{fig8}.}\label{fig7}
\end{figure}

\begin{figure} [h]
\resizebox{\hsize}{!}{\includegraphics{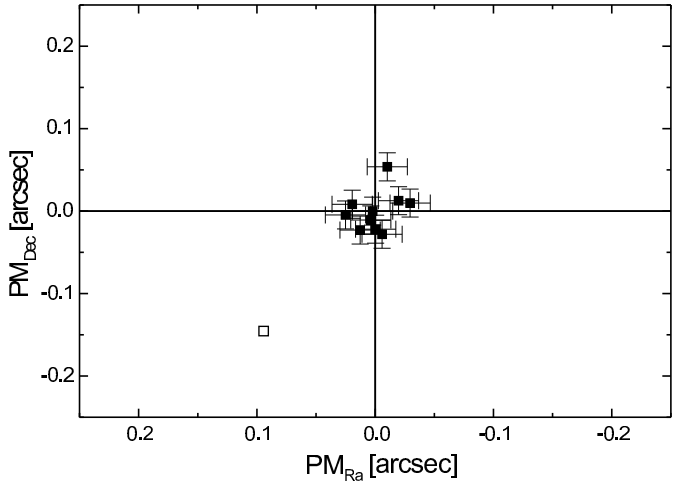}}\caption{The measured proper motions of
all detected companion candidates within the UFTI field of view around the exoplanet host star
HD\,40979. The expected proper motion of a co-moving companion is shown by a white small box that
is derived with the given epoch difference and the Hipparcos astrometry of the exoplanet host star.
All detected objects turn out to be slow-moving background stars.}\label{fig8}
\end{figure}

We observed HD\,40979 the first time in November 2002 with a second-epoch observation one year
later in October 2003. The first-epoch H-band image is shown in Fig.\,\ref{fig7}. By comparing both
UFTI images, we obtain the proper motions of all companions-candidates ($S/N$\,$>$\,10), which are
detected in both UFTI images. The expected proper and parallactic motion of the exoplanet host star
between our first and second-epoch imaging is well-known from precise Hipparcos astrometry
($\mu_{\alpha}cos(\delta)$\,=\,95.05$\pm$0.87\,mas/yr,
$\mu_{\delta}$\,=\,$-$152.23$\pm$0.47\,mas/yr and $\pi$\,=\,30.00$\pm$0.82\,mas). The measured
proper motions of all detected companion candidates in our UFTI images are shown in
Fig.\,\ref{fig8}. All detected objects around HD\,40979 emerge as slowly moving stars that are
unrelated to the exoplanet host star, most probably located far away in the background.

\begin{figure} [h]
\resizebox{\hsize}{!}{\includegraphics{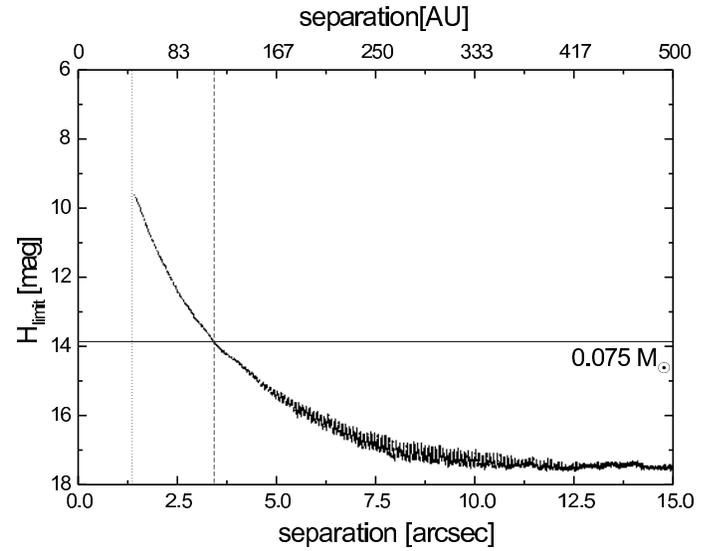}}\caption{The detection limit of our UFTI
images of HD\,40979 for a range of separations given in arcsec at the bottom and as projected
separation in AU at the top. At $\sim$1.4\,arcsec (47\,AU), saturation occurs (dotted line). At an
assumed system age of 1\,Gyr all stellar companions (m\,$\ge$\,0.075\,$M_{\sun}$) are detectable
beyond 3.4\,arcsec ($\sim$110\,AU), see dashed line.}\label{fig9}
\end{figure}

The detection limit of our UFTI observations is shown in Fig.\,\ref{fig9}. Objects within a radius
of $\sim$53\,arcsec ($\sim$1800\,AU) around the exoplanet host star are detected in both UFTI
images. In the background-limited region at separations larger than $\sim$10\,arcsec (350\,AU), a
limiting magnitude of H\,=\,17.5\,mag is reached. According to the \citet{baraffe2003} models at an
assumed system age of 1\,Gyr, this limiting magnitude enables the detection of substellar
companions with masses, down to 0.032\,$M_{\sun}$. All stellar companions
(m\,$>$\,0.075\,$M_{\sun}$) are detectable beyond 3.4\,arcsec ($\sim$110\,AU).

\begin{figure} [h]
\resizebox{\hsize}{!}{\includegraphics{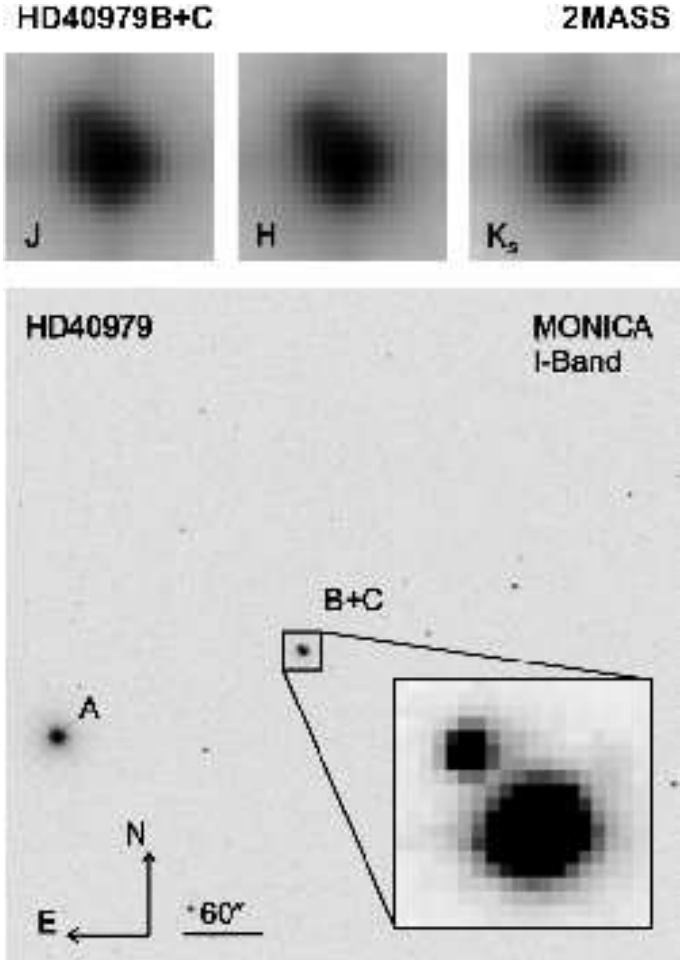}}\caption{The 2MASS J-, H-, and $K_{S}$-band
images (top panel) of HD\,40979\,B and one of our first-epoch MONICA images (bottom panel), which
shows all components of the triple-star system HD\,40979. The pattern of HD\,40979\,B in the three
2MASS images appears elongated, which is a first hint that there might be an additional faint
object located only a few arcsec northeast of HD\,40979\,B. This object is clearly separated from
HD\,40979\,B in our MONICA images.}\label{fig10}
\end{figure}

Although we did not discover any companion to HD\,40979, \cite{halbwachs1986} reported that this
star has a known co-moving companion, with a wide angular separation of 192.5\,arcsec
($\sim$6400\,AU). The proper motion of the co-moving companion HD\,40979\,B is listed in the UCAC2
catalogue ($\mu_{\alpha}cos(\delta)$\,=\,93.8$\pm$0.7\,mas/yr,
$\mu_{\delta}$\,=\,$-$154.0$\pm$0.7\,mas/yr). Due to the large angular separation between the two
stars, the co-moving companion lies outside the UFTI field of view, but is detected in the 2MASS
image, together with the exoplanet host star. The upper panel of Fig.\,\ref{fig10} shows the J, H,
and K$_{S}$ 2MASS images of the co-moving companion HD\,40979\,B. Note that the companion appears
elongated in the northeast direction, which is different from the diffraction patterns of the
nearby stars. The elongated diffraction pattern might indicate that there is an unresolved faint
object located only a few arcsec northeast of the wide co-moving companion HD\,40979\,B.

In order to confirm this result, we observed HD\,40979\,B with the 0.8\,m telescope of the Munich
LMU University Observatory, located on Mount Wendelstein (Bavaria, south Germany). This telescope
is equipped with the CCD-camera MONICA (MOnochromatic Image CAmera), a 1024\,$\times$\,1024 SiTe
CCD-array (Textronik TK 1024) with a pixel scale of $\sim$490\,mas per pixel and a large field of
view of 8.4\,arcmin\,$\times$\,8.4\,arcmin. We observed the common-proper-motion pair for the first
time in November 2004 in the I-band (see Fig.\,\ref{fig10}). Due to the large MONICA field of view
both binary components HD\,40979\,A and B are detected on the CCD. In order to avoid saturation, we
used an exposure time of only 1\,s. In total nine I-band images were taken and one of these images
is shown in the lower pattern of Fig.\,\ref{fig10}.

As already indicated by the elongated diffraction pattern of HD\,40979\,B in the 2MASS images,
there is indeed a faint object that is located northeast of HD\,40979\,B. Because of their small
angular separation ($\sim$3.9\,arcsec), the two stars are not resolved in 2MASS images, whose pixel
scale is 1\,arcsec per pixel, but clearly stand out in our MONICA I-band images. The faint object
might be either an additional companion of the system or just an ordinary background star. To find
out if the faint star is really associated with HD\,40979\,B, we carried out in July 2005 follow-up
second-epoch observations with the same image preset.

\begin{figure} [h]
\resizebox{\hsize}{!}{\includegraphics{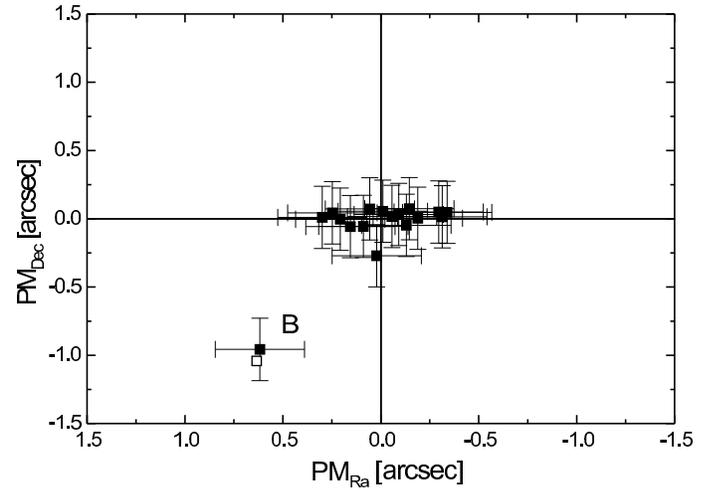}}\caption{The derived proper motions of
all objects detected around the exoplanet host star HD\,40979\,A. Comparison is between our second
epoch MONICA images taken in July 2005 and the 2MASS images from 1998. The expected proper motion
of a companion who shares the proper motion of the exoplanet host star object is indicated as a
small white square in the plot.}\label{fig11}
\end{figure}

In a first step, we compared our second-epoch MONICA images with the 2MASS images and determined
the proper motions of all objects that appear in the two sets of images (see Fig.\,\ref{fig11}).
HD\,40979\,B and the exoplanet host star HD\,40979\,A clearly form a common-proper-motion pair,
whereas all other detected objects emerge as unrelated slow-moving objects. The separations and
position angles of HD\,40979\,B relative to its primary star for all three observing epochs are
listed in Table\,\ref{table5}. They are constant within the astrometric uncertainties as expected
for a common-proper-motion pair. We also list the expected variation in separation and position
angle if HD\,40979\,B is a non-moving background source, using the proper and parallactic motion of
the exoplanet host star (see the columns $sep$$_{~if~bg}$ and $PA$$_{~if~bg}$).

\begin{table}[ht]
\caption{The separations and position angles of HD\,40979\,B relative to the exoplanet host star
HD\,40979\,A.}
\begin{center}
\begin{tabular}{c|c|c|c}
\hline\hline
HD\,40979\,B& epoch & $sep$[arcsec] & $sep$$_{~if~bg}$ [arcsec]\\
\hline
            & 2MASS 11/98          & 192.489$\pm$0.085 & ---\\
            & Wend   11/04         & 192.586$\pm$0.070 & 193.329\\
            & Wend   07/05         & 192.351$\pm$0.114 & 193.432\\
\hline
HD\,40979\,B& epoch             & PA [$^{\circ}$]   & PA$_{~if~bg}$ [$^{\circ}$]\\
\hline
            & 2MASS 11/98        & 289.230$\pm$0.030   & ---\\
            & Wend 11/04         & 289.275$\pm$0.037   & 289.432\\
            & Wend 07/05         & 289.256$\pm$0.063   & 289.459\\
\hline\hline
\end{tabular}
\end{center}\label{table5}
\end{table}

The newly detected faint companion candidate, located northeast of HD\,40979\,B is well-resolved in
both MONICA images. We measured its separation and position angle relative to HD\,40979\,B in both
MONICA images and summarize all results in Table\,\ref{table7}. If the companion candidate is a
non-moving background object, we expect a change in the separation and position angle due to the
proper and parallactic motion of HD\,40979\,B between the observations. The expected variation in
both parameters is given in Table\,\ref{table7}. The position angle has to change significantly in
the case of a non-moving background source (see $sep_{~if~bg}$ and $PA_{~if~bg}$) but the position
angle and separation of the faint object remain constant within their uncertainties in both MONICA
images. Hence, this faint object is a co-moving companion of HD\,40979\,B and is therefore denoted
as HD\,40979\,C. We therefore conclude that the wide binary HD\,40979 is a triple-star system, with
the exoplanet host star HD\,40979\,A as primary and a widely separated secondary pair whose
components are separated by 129\,AU (projected separation).

\begin{table}[ht]
\caption{The separations and position angles of HD\,40979\,C relative to HD\,40979\,B.}
\begin{center}
\begin{tabular}{c|c|c|c}
\hline\hline
HD\,40979\,C& epoch & sep[arcsec] & sep$_{~if~bg}$ [arcsec]\\
\hline
            & Wen 11/04         & 3.857$\pm$0.014 & ---\\
            & Wen 07/05         & 3.877$\pm$0.013 & 3.913\\
\hline
HD\,40979\,C& epoch             & PA [$^{\circ}$]   & PA$_{~if~bg}$ [$^{\circ}$]\\
\hline
            & Wen 11/04         & 37.811$\pm$0.215  & ---\\
            & Wen 07/05         & 37.969$\pm$0.178  & 35.972\\
\hline\hline
\end{tabular}
\end{center}\label{table7}
\end{table}

We measured the apparent I-band magnitudes of HD\,40979\,B and C in our first and second-epoch
MONICA images and obtained $I(B)$\,=\,7.98$\pm$0.03\,mag and $I(C)$\,=\,11.3$\pm$0.1\,mag. With the
distance of the exoplanet host star, we find that the absolute I-band magnitudes of both companions
are M$_{I}(B)$\,=\,5.37$\pm$0.07\,mag and M$_{I}(C)$\,=\,8.67$\pm$0.12\,mag. Converting their
I-band magnitudes to masses with the \citet{baraffe1998} models, we obtain, for a system age of
1\,Gyr, a mass of 0.833$\pm$0.011\,$M_{\sun}$ for HD\,40979\,B and 0.380$\pm$0.025\,$M_{\sun}$ for
HD\,40979\,C, respectively. These mass-estimates do not change significantly for system ages
between 1 and 10\,Gyr.

\section{Spectroscopy}

In this section we present our spectroscopic observations of the two co-moving companions
GJ\,3021\,B and HD\,27442\,B, which confirm their companionship to the exoplanet host stars.

\subsection{Near infrared spectroscopy of GJ\,3021\,B and HD\,27442\,B}

We obtained near infrared H- and K-band spectra of GJ\,3021\,B and HD\,27442\,B in October and
December 2004 at Paranal Observatory. We used the Infrared Spectrometer And Array Camera (ISAAC),
which is mounted at one of the Nasmyth foci of UT1 (Antu). ISAAC is equipped with a
1024\,$\times$\,1024 Hawaii Rockwell IR detector ($\sim$0.148\,arcsec per pixel) for imaging and
spectroscopy in the near infrared J-, H-, and K-bands. We used the low-resolution grism, together
with the order sorting filters SH (H-band) and SK (K-band), as well as the 1\,arcsec slit that
yields a resolving power $\lambda/\Delta\lambda$ of $\sim$500 in H and $\sim$450 in the K-band with
dispersions of 4.8\,\AA~per pixel and 7.2\,\AA~per pixel, respectively.

Background subtraction was obtained by a 45\,arcsec nodding between two positions along the slit.
In addition we applied a 5\,arcsec random jitter around the two nodding positions to avoid the
individual pixels always seeing the same part of the sky. We obtained 16 spectra per target each
with an integration time of 60\,s. All spectra were flat-fielded using internal lampflats and
wavelength-calibrated with Xenon and Argon arclamps. After background subtraction, flat-fielding
and wavelength calibration, all done with standard \textsl{IRAF} routines, the individual spectra
were finally averaged.

\begin{figure} [h]
\resizebox{\hsize}{!}{\includegraphics{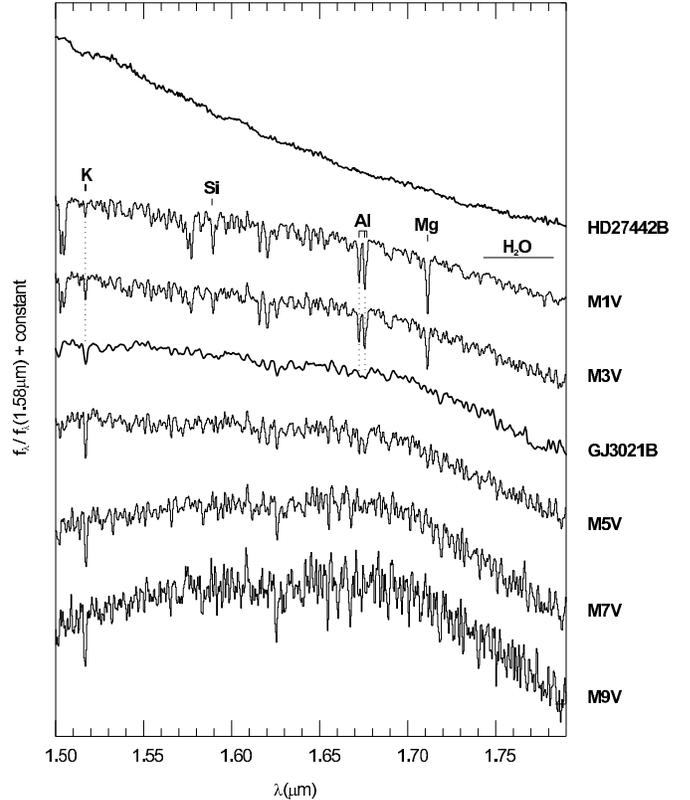}}\caption{Our flux-calibrated H-band ISAAC spectra
of the co-moving companions GJ\,3021\,B and HD\,27422\,B. In addition we show M1V to M9V comparison
spectra from \citet{cushing2005}.}\label{fig12}
\end{figure}

\begin{figure} [h]
\resizebox{\hsize}{!}{\includegraphics{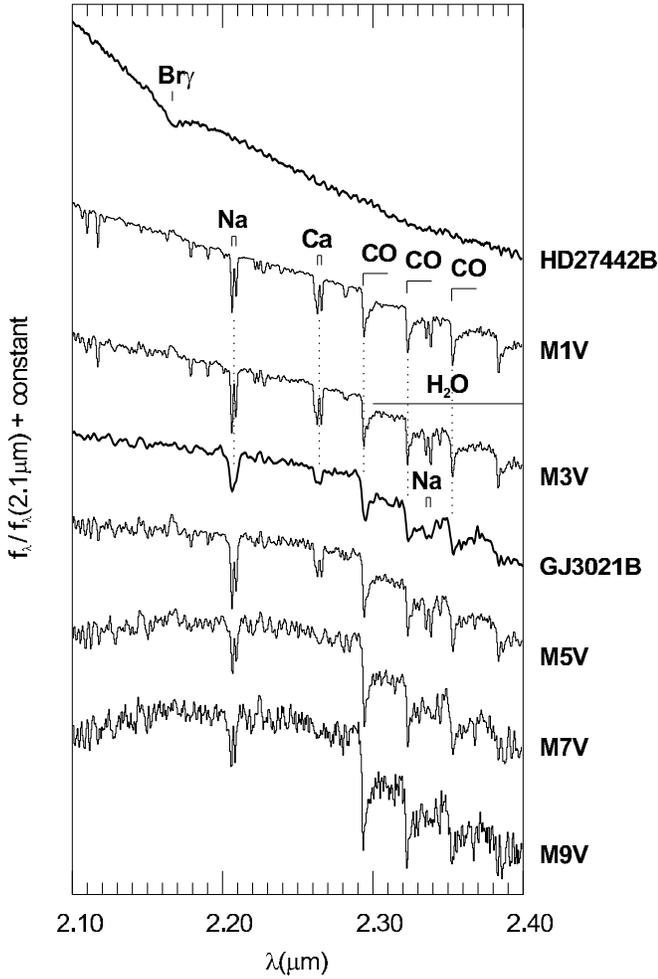}}\caption{The flux-calibrated K-band ISAAC spectra
of GJ\,3021\,B and HD\,27442\,B, together with M1V to M9V comparison spectra from
\citet{cushing2005}.}\label{fig13}
\end{figure}

The reduced spectra of GJ\,3021\,B and HD\,27442\,B were calibrated with the telluric standard
stars HD\,50491 (B4V) and HD\,48215 (B5V), respectively. The spectra were always taken directly
after the science spectra with a maximal airmass difference of only $\sim$0.2\,dex. We determined
the response function of the spectrograph with the spectra of the spectroscopic standard stars and
reference spectra from \citet{pickles1998}. The H- and K-band spectra of GJ\,3021\,B and
HD\,27442\,B are shown in Figs.\,\ref{fig12} and \ref{fig13}, together with comparison spectra of
dwarfs with spectral types between M1 and M9 from \citet{cushing2005}.

The most dominant absorption feature in the H-band spectrum of GJ\,3021\,B is the potassium line at
1.52\,$\mu$m. The Mg absorption line at 1.71\,$\mu$m was not detected. Instead, we find the faint
absorption features of Al at 1.67\,$\mu$m. The shape of the continuum is consistent with a spectral
type between M3 and M5.

In the K-band spectrum of GJ\,3021\,B, we find atomic absorbtion features of sodium at 2.21\,$\mu$m
and 2.33\,$\mu$m, calcium at 2.26\,$\mu$m, and the series of the first-overtone band heads of CO
which extend redward of 2.29\,$\mu$m. The strength of the molecular features compared with detected
atomic lines is a sensitive indicator of the luminosity class. For equal spectral types the
molecular absorption features are more prominent in stars with lower surface gravities (e.g.
giants) than in main sequence stars. However, the detected atomic absorbtion features of Na and Ca
in the K-band spectrum of GJ\,3021\,B are comparable to the CO molecular band heads; i.e.
GJ\,3021\,B is a dwarf with a spectral type M3V to M5V.

Our photometric and astrometric results corroborate the results of \cite{chauvin2006}, which show
that GJ\,3021\,B is a co-moving companion with a photometry compatible with what is expected for a
mid-M dwarf. Our ISAAC spectroscopy unambiguously confirmed this approximation and finally confined
the spectral type of the companion to M4+-1V.

The H- and K-band spectra of HD\,27442\,B are clearly different from the comparison M-dwarf spectra
shown in Figs.\,\ref{fig12} and \ref{fig13}. The H- and K-band continua are even steeper than the
earliest comparison dwarf M1, indicating a high effective temperature hotter than 3700\,K (M1). The
H-band spectrum is featureless, whereas we find the $Br_{\gamma}$ line at 2.16\,$\mu$m in the
K-band. This absorption feature is most prominent in B to late F spectral types and it weakens
towards spectral types of early K \citep[$T_{eff}>5000$\,K, see][]{wallace1997}.

Our H- and K-band spectroscopy clearly confirms that HD\,27442\,B is not a cool low-mass M-dwarf
companion. Instead, the lower limit of its effective temperature is consistent with the white dwarf
hypothesis, which was concluded in Sect.\,2 from photometric data alone. Furthermore, the detected
$Br_{\gamma}$ line indicates that HD\,27442\,B exhibits a hydrogen atmospheric layer, and we expect
to find more typical Hydrogen lines in the spectrum of this companion. However, no other lines of
the Bracket series are visible in the H-band. To confirm the presence of Hydrogen in the atmosphere
of HD\,27442\,B, we also obtained optical spectra of this companion, reported in the next section.

\subsection{Optical spectroscopy of HD\,27442\,B}

We observed HD\,27442\,B in December 2004 at Paranal Observatory with the second visual and near-UV
FOcal Reducer and low dispersion Spectrograph (FORS2), which is installed on the Cassegrain focus
of UT1 (Antu). FORS2 is equipped with a mosaic of two MIT CCD detectors, each a
4096\,$\times$\,2048 pixel array with a pixel scale of 0.126\,arcsec per pixel. For long-slit
spectroscopy, we used the 300I+21 grism combined with the order-sorting filter OG590. With the
1\,arcsec slit and the standard resolution collimator, we obtained a resolving power of
$\lambda/\Delta\lambda$\,=\,660 at the central wavelength 0.86\,$\mu$m. The CCD arrays were
operated with a 2x2 pixel binning and a dispersion of 3.22\,\AA~per pixel.

We took 8 FORS spectra of HD\,27442\,B, each with an integration time of 200\,s. Between the
exposures the telescope was nodded along the slit to avoid always taking the spectrum by the same
pixel (bad pixel correction). The bias level was removed, from each spectrum, which was then
flat-fielded with a screen flat-field image. Finally all spectra were wavelength-calibrated with a
composite spectrum of a helium-argon lamp. The slit was orientated in such a way that the companion
and a small part of the primary PSF were always located on the slit. In this way we simultaneously
obtained spectra of the companion and of the primary whose spectral type is well-known (K2IV) and
can be used for flux calibration. We derived the response function of the spectrograph using the
primary spectrum and flux-calibrated reference spectra from \citet{danks1994}\footnote{telluric
absorption features not removed in these spectra}.

\begin{figure} [h]
\resizebox{\hsize}{!}{\includegraphics{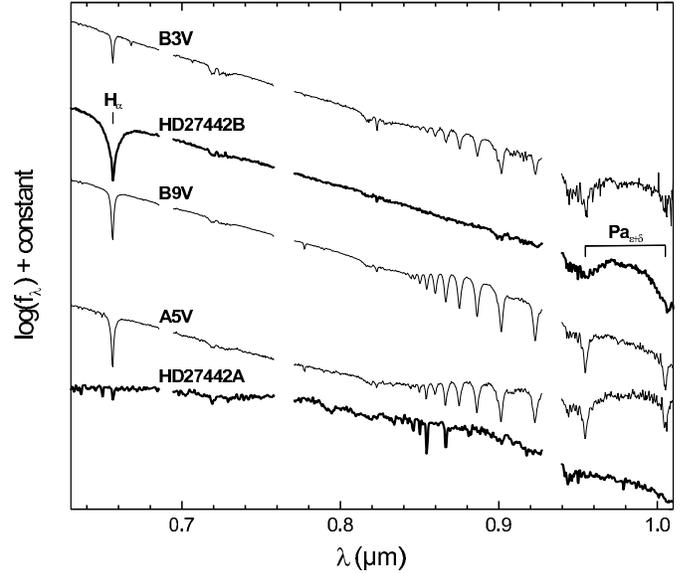}}\caption{The FORS2 spectrum of the exoplanet
host star HD\,27442\,A (K2IV) and its co-moving companion HD\,27442\,B, together with reference
spectra with spectral types from mid B to mid A from \citet{danks1994}. Prominent telluric
absorption features are blanked in all spectra. }\label{fig14}
\end{figure}

We show the flux-calibrated spectra of the exoplanet host star HD\,27442\,A and its co-moving
companion HD\,27442\,B in Fig.\,\ref{fig14}, together with comparison spectra of mid B to mid A
dwarfs from \citet{danks1994}. The most striking telluric absorption features are blanked in all
spectra. The spectrum of HD\,27442\,B is clearly different from the primary spectrum, and it
exhibits a much steeper continuum. This continuum is comparable to the continua of mid to late B
dwarfs with spectral types in the range B3 (11000K) and B9 (19000K). Furthermore, we detected
hydrogen absorption lines, which are all much broader than the absorption lines of the comparison
dwarfs. The line broadening is attributed to a high surface gravity and is typical of the
absorption lines of white dwarfs. The most prominent hydrogen absorption feature is the Balmer line
$H_{\alpha}$ at 0.66\,$\mu$m. We also find two broadened lines of the Paschen series
--- $Pa_{\epsilon}$ at 0.95\,$\mu$m and $Pa_{\delta}$ at 1.01\,$\mu$m.

We detected in total four Hydrogen absorption features in the visible and near infrared spectral
range. Our photometric and astrometric results corroborate the results of \cite{chauvin2006}, which
show that HD\,27442\,B is a co-moving companion with a visible and infrared photometry compatible
with that expected for a white dwarf. Finally, our ISAAC and FORS2 spectroscopic observations
unambiguously confirm the white dwarf nature of HD\,27442\,B.

The companionship of HD\,40979\,B+C has so far been proven only with astrometry; i.e. the
companions clearly share the proper motion of the exoplanet host star HD\,40979\,A. With the
derived absolute magnitudes of the B and C components, we can derive the expected V$-$K colors of
these two stars using the \cite{baraffe1998} models. We get V$-$H\,$\sim$2.4 for HD\,40979\,B and
V$-$H\,$\sim$4.2 for HD\,40979\,C. According to the color-spectral type conversion of
\citet{kenyon1995}, these colors correspond to spectral types of K3 and M3, respectively. We also
plan to take follow-up spectra to confirm these results.

\begin{table*}[pht]
\caption{Double and triple stars among the exoplanet host stars.}

\begin{center}
\resizebox{\hsize}{!}{\begin{tabular}[h!]{l|c|c|c|c|c|c|c|c|c|c|c|c|c} \hline\hline
\multicolumn{14}{c}{Class-1 systems: a,e assumed}\\
\hline\hline
planet host star& mass$_{Host}$   & mass$_{Comp}$ & a$_{Bin}$ & e$_{Bin}$ & $\mu$     & a$_{c}$ & a$_{P}$ & P      & e    & msin(i)     & K     & ID & ref\\
host star&[$M_{\sun}$]&[$M_{\sun}$]             & [AU]      &           &           & [AU]    & [AU]    & [days] &      & [$M_{Jup}$] & [m/s] &    &    \\
\hline 55\,Cnc\,A   &   0.87    &   0.27        &   1062    &   0.5     &   0.233   &   164   & 0.04    & 3      & 0.17 & 0.05        & 6.7   & e  & McArthur 2004\\
                    &           &               &           &           &           &         & 0.12    & 15     & 0.02 & 0.78        & 67.4  & b  & McArthur 2004\\
                    &           &               &           &           &           &         & 0.24    & 44     & 0.44 & 0.22        & 13.0  & c  & McArthur 2004\\
                    &           &               &           &           &           &         & 5.26    & 4517   & 0.33 & 3.91        & 49.8  & d  & McArthur 2004\\
83\,Leo\,B          &   0.78    &   0.9         &   502     &   0.5     &   0.537   &   57    & 0.12    & 17     & 0.05 & 0.11        & 10.5  & b  & Marcy 2005\\
GJ\,3021\,A         &   0.99    &   0.13        &   68      &   0.5     &   0.112   &   12    & 0.49    & 134    & 0.51 & 3.37        & 167   & b  & Naef 2001a\\
GJ\,777\,A          &   0.96    &   0.2         &   2841    &   0.5     &   0.172   &   462   & 0.13    & 17     & 0.01 & 0.06        & 4.6   & c  & Vogt 2005\\
                    &           &               &           &           &           &         & 3.92    & 2891   & 0.36 & 1.50        & 23.5  & b  & Vogt 2005\\
Gl\,86\,A           &   0.70    &   0.55        &   21      &   0.5     &   0.440   &   3     & 0.11    & 16     & 0.05 & 4           & 380   & b  & Queloz 2000\\
HD\,109749\,A       &   1.04    &   0.18        &   489     &   0.5     &   0.144   &   81    & 0.06    & 5      & 0.01 & 0.28        & 28.8  & b  & Fischer 2006\\
HD\,114729\,A       &   0.97    &   0.25        &   282     &   0.5     &   0.207   &   44    & 2.08    & 1135   & 0.32 & 0.84        & 18    & b  & Butler 2003\\
HD\,114762\,A       &   0.81    &   0.09        &   134     &   0.5     &   0.097   &   23    & 0.34    & 84     & 0.35 & 9           & 590   & b  & Mazeh 1996\\
HD\,142\,A          &   1.28    &   0.56        &   104     &   0.5     &   0.305   &   15    & 1.0     & 339    & 0.37 & 1.03        & 30    & b  & Tinney 2002\\
HD\,142022\,A       &   0.99    &   0.66        &   731     &   0.5     &   0.401   &   96    & 3.03    & 1928   & 0.53 & 5.1         & 92    & b  & Eggenberger 2006\\
HD\,16141\,A        &   1.05    &   0.29        &   223     &   0.5     &   0.214   &   35    & 0.35    & 76     & 0.28 & 0.22        & 11    & b  & Marcy 2000\\
HD\,195019\,A       &   1.06    &   0.67        &   128     &   0.5     &   0.387   &   17    & 0.13    & 18     & 0.02 & 3.47        & 272   & b  & Vogt 2000\\
HD\,196050\,A       &   1.15    &   0.36        &   510     &   0.5     &   0.240   &   78    & 2.4     & 1300   & 0.19 & 2.8         & 49    & b  & Jones 2002\\
HD\,213240\,A       &   1.22    &   0.15        &   3899    &   0.5     &   0.107   &   668   & 2.03    & 951    & 0.45 & 4.5         & 91    & b  & Santos 2001\\
HD\,222582\,A       &   1.02    &   0.38        &   4596    &   0.5     &   0.272   &   684   & 1.35    & 576    & 0.71 & 5.29        & 184   & b  & Vogt 2000\\
HD\,27442\,A        &   0.98    &   0.58        &   236     &   0.5     &   0.372   &   32    & 1.16    & 415    & 0.06 & 1.35        & 33    & b  & Butler 2001\\
HD\,46375\,A        &   0.82    &   0.58        &   346     &   0.5     &   0.413   &   45    & 0.04    & 3      & 0.04 & 0.25        & 35    & b  & Marcy 2000\\
HD\,75289\,A        &   1.23    &   0.14        &   621     &   0.5     &   0.099   &   107   & 0.05    & 4      & 0.02 & 1.15        & 54    & b  & Udry 2000\\
HD\,80606\,A        &   1.04    &   0.90        &   1200    &   0.5     &   0.466   &   147   & 0.47    & 112    & 0.93 & 3.90        & 411   & b  & Naef 2001\\
HD\,89744\,A        &   1.53    &   0.08        &   2456    &   0.5     &   0.049   &   440   & 0.88    & 256    & 0.70 & 7.2         & 257   & b  & Korzennik 2000\\
$\upsilon$\,And\,A  &   1.30    &   0.19        &   749     &   0.5     &   0.125   &   127   & 0.06    & 5      & 0.04 & 0.72        & 74    & b  & Butler 1999\\
                    &           &               &           &           &           &         & 0.83    & 242    & 0.23 & 1.98        & 56    & c  & Butler 1999\\
                    &           &               &           &           &           &         & 2.50    & 1269   & 0.36 & 4.11        & 70    & d  & Butler 1999\\
HD\,40979\,A        &   1.21    &   1.21 (S+S)  &   6416    &   0.5     &   0.500   &   756   & 0.83    & 263    & 0.25 & 3.28        & 101   & b  & Fischer 2003\\
HD\,41004\,A        &   0.91    &   0.59 (S+B)  &   23      &   0.5     &   0.395   &   3     & 1.7     & 963    & 0.74 & 2.54        & 99    & b  & Zucker 2004\\
HD\,178911\,B       &   0.98    &   1.88 (S+S)  &   784     &   0.5     &   0.657   &   76    & 0.32    & 72     & 0.12 & 6.29        & 339   & b  & Zucker 2002a\\
\hline\hline
\end{tabular}}
\end{center}
\begin{center}
\resizebox{\hsize}{!}{\begin{tabular}[h!]{l|c|c|c|c|c|c|c|c|c|c|c|c|c} \hline\hline
\multicolumn{14}{c}{Class-2 systems: a,e known}\\
\hline\hline
planet host star& mass$_{Host}$   & mass$_{Comp}$ & a$_{Bin}$ & e$_{Bin}$ & $\mu$     & a$_{c}$ & a$_{P}$ & P      & e    & msin(i)     & K     & ID & ref\\
host star&[$M_{\sun}$]&[$M_{\sun}$]             & [AU]      &           &           & [AU]    & [AU]    & [days] &      & [$M_{Jup}$] & [m/s] &    &    \\
\hline
$\gamma$\,Cep\,A   &   1.59    &   0.41        &   19    &   0.361   &   0.207   &   4     & 2.13    & 906    & 0.12 & 1.7         & 28    & b  & Hatzes 2003\\
HD\,19994\,A        &   1.37    &   0.62        &   151     &   0.26    &   0.310   &   36    & 1.42    & 536    & 0.30 & 1.68        & 36    & b  & Mayor 2004\\
$\tau$\,Boo\,A      &   1.33    &   0.44        &   98      &   0.4189  &   0.247   &   18    & 0.05    & 3      & 0.02 & 3.87        & 469   & b  & Butler 1997\\
HD\,188753\,A       &   1.06    &   1.63 (S+S)  &   12    &   0.5     &   0.606   &   1     & 0.05    & 3      & 0    & 1.14        & 149   & b  & Konacki 2005\\
16\,Cyg\,B          &   0.99    &   1.18 (S+S)  &   755     &   0.863   &   0.543   &   15    & 1.6     & 801    & 0.63 & 1.5         & 44    & b  & Cochran 1997\\
\hline\hline
\end{tabular}}
\end{center}

\begin{center}
\tiny{
\begin{tabular}{l l}
\underline{Remarks:}&\\
\\
55\,Cnc\,A:& separation and mass of B component from \citet{mugrauer2006}, this also holds for HD\,80606\,A and HD\,46375\,A\\
83\,Leo\,B: & fainter B component is the exoplanet host star whose mass is estimated by \citet{marcy2005}\\
Gl\,86\,A:& B is a white dwarf; separation and mass estimated by \citet{mugrauer2005a}\\
HD\,142\,A:& mass of B component determined with our SofI H-band photometry\\
HD\,109749\,A: & mass of primary from \citet{allende1999}\\
HD\,114729\,A: & separation and mass of B component from \citet{mugrauer2005b}, this also holds for HD\,196050, HD\,213240 and HD\,16141\\
HD\,114762\,A: & mass of B component determined with 2MASS K-band photometry (A+B unresolved) and $\Delta$K$_{AB}$ from \citet{patience2002}\\
HD\,142022\,A & mass of primary from \citet{eggenberger2006}\\
HD\,27442\,A: & primary mass from \citet{randich1999}\\
HD\,75289\,A:& separation and mass of companion from \citet{mugrauer2004a}\\
HD\,89744\,A:& separation and mass of companion from \citet{mugrauer2004b}\\
HD\,40979\,A: & triple system A/B+C; mass of B and C component are derived with I-band MONICA photometry (see section 2)\\
HD\,41004\,A:& triple system A/B+C with a brown-dwarf companion (C); the separation between A and B+C is measured by Hipparcos; primary mass from\\
& \citet{allende1999} and mass of B comp. is derived with Hipparcos photometry; mass of brown-dwarf companion estimated with\\
& radial velocity data from \citet{zucker2004}\\
HD\,19994\,A: & mass of B comp. derived with 2MASS K-band photometry (A+B unresolved) and $\Delta$K$_{AB}$\,=\,3.0$\pm$0.1\,mag measured by us with NACO/VLT\\
& in run 073.C-0370(A)\\
$\tau$\,Boo\,A: & mass of B component determined with 2MASS K-band photometry (A+B unresolved) and $\Delta$K$_{AB}$ from \citet{patience2002}\\
HD\,188753\,A: & triple system A/B+C, all masses from \citet{konacki2005b}\\
HD\,178911\,B: & triple system A+C/B, masses of A and C component from \citet{tokovinin2000}\\
16\,Cyg\,B & triple system A+C/B, mass of C comp. is determined with 2MASS H-band photometry (A+C unresolved) and $\Delta$H$_{AC}$ from \citet{patience2002};\\
$\gamma$\,Cep: & orbital solution and mass determination of B comp. with radial velocity data from \citet{hatzes2003}; mass of A comp. from \citet{fuhrmann2004}\\
\end{tabular}}
\end{center}
\label{table8}
\end{table*}

\nocite{butler1997}\nocite{butler1999}\nocite{butler2001}\nocite{butler2003}\nocite{cochran1997}\nocite{eggenberger2006}
\nocite{fischer2003}\nocite{fischer2006}\nocite{hatzes2003}\nocite{jones2002}\nocite{konacki2005b}\nocite{korzennik2000}
\nocite{mcarthur2004}\nocite{marcy2000}\nocite{marcy2005}\nocite{mazeh1996}\nocite{mayor2004}\nocite{naef2001a}
\nocite{naef2001b}\nocite{queloz2000}\nocite{santos2001}\nocite{tinney2002}\nocite{udry2000}\nocite{vogt2000}
\nocite{vogt2005}\nocite{zucker2002a}\nocite{zucker2004}

\section{Exoplanet host stars in multiple-star systems}

\cite{eggenberger2004} have already compiled a list of 14 planet host stellar systems, and until
end of 2005, two other systems have been reported --- HD\,142022 \citep{eggenberger2006} and
HD\,188753 \citep{konacki2005b}.

In addition to our multiplicity study of the exoplanet host stars, we also identified other
exoplanet host stars listed as binaries in the WDS catalogue
--- 83\,Leo\,B, HD\,109749, HD\,222582, and HD\,142. The proper motions of both components of
83\,Leo\,B, HD\,109749, and HD\,222582 are listed in either the Hipparcos or UCAC2 astrometric
catalogues, and the apparent 2MASS photometry of the co-moving companions is consistent with
low-mass stars located at the distances of the exoplanet host stars. Thus, the multiplicity of
these systems is confirmed by both astrometry and photometry.

The third star, HD\,142, is a close binary with separation of \,$\sim$5\,arcsec) and is listed in
the WDS with 5 astrometric measurements. We observed HD\,142 with SofI in December 2002 and again
in June 2003.  According to the Hipparcos astrometry, the expected motion of the primary between
the two observations is $PM_{Dec}$\,=\,357$\pm$1\,mas and $PM_{Dec}$\,=\,5$\pm$1\,mas. With both
SofI images, we obtained the proper motion of the WDS companion $PM_{Ra}$\,=\,345$\pm$30\,mas and
$PM_{Dec}$\,=\,43$\pm$30\,mas, which confirms the common proper motion of the pair. The photometry
of the B component is contaminated in the 2MASS images by the flux of the nearby bright exoplanet
host star; hence, the 2MASS photometry is very uncertain (H\,=\,5.246\,mag, no magnitude error
given here). However, the co-moving companion is well-separated from the primary in our SofI H-band
images. We determined its apparent H-band magnitude H\,=\,7.613$\pm$0.031\,mag. With the Hipparcos
parallax of the primary (39.00$\pm$0.64\,mas), this yields the absolute magnitude of the companion
$M_{H}$\,=\,5.568$\pm$0.047\,mag. According to the magnitude-mass relation of the evolutionary
\citet{baraffe1998} models, this magnitude corresponds to a mass of 0.56\,$M_{\sun}$ assuming a
system age of 5\,Gyr (see Table\,\ref{table8}).

The updated list of planet host multiple-star systems is presented in Table\,\ref{table8}. We list
the mass of the exoplanet host stars and of the secondaries. If the secondary itself is a binary
system, we always show the total mass of the pair and add S+S in the case of two stars or S+B in
the case of a star with a brown-dwarf companion. We derive the ratio between the companion mass and
the total system-mass ($\mu$), as well as the critical semi-major axes (a$_{c}$). We also list the
orbital properties of all exoplanets detected in these stellar systems.

We consider here only systems whose exoplanets ($msin(i)$\,$<$\,13\,$M_{Jup}$) are published in
refereed papers until the end of 2005. We excluded two systems from the table
--- the WDS binary HD\,11964, and the triple system HD\,219449\,A+BC, because the exoplanets in
these stellar systems were both announced, but not published in a refereed paper. In total, the
table includes 29 systems, 24 binaries, and 5 triple systems --- HD\,188753, 16\,Cyg, HD\,178911,
HD\,40979, and HD\,41004.

The whole sample can be subdivided into two classes of systems. Class-1 systems are confirmed
common-proper-motion pairs whose projected separations are known from astrometric measures but
their orbital parameters (e.g. semi-major axes, eccentricity) are unknown. For these systems we
always used the projected separations as an estimate of the semi-major axes and assumed an orbital
eccentricity of e\,=\,0.5, which is a good estimate for wide multiple-star systems (see e.g.
eccentricity distribution of wide binaries from Soederhjelm 1999 \nocite{soederhjelm1999}). For 5
systems, only the orbital parameters are known from either astrometric \citep{hartkopfonline} or
radial-velocity measurements \citep{hatzes2003}. These special systems are denoted as Class-2
systems.

In the majority of cases (exceptions are summarized under remarks in Table\,\ref{table8}), we
derived the masses of the companions using their apparent 2MASS infrared magnitudes, the Hipparcos
parallaxes of the exoplanet host stars, and the magnitude-mass conversion of the 5\,Gyr
\citet{baraffe1998} models. For the masses of the exoplanet host stars, we used the mass-estimates
from \citet{santos2004}, if available (for exceptions see remarks in Table\,\ref{table8}).

\begin{figure} [h!]
\resizebox{\hsize}{!}{\includegraphics{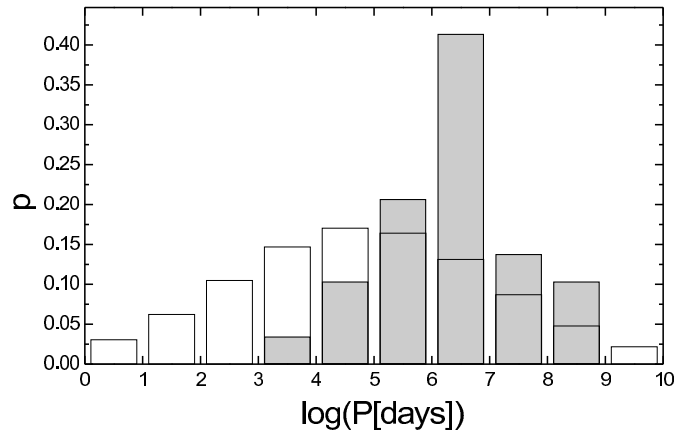}}\caption{The relative frequencies of the
approximated orbital periods of planet host binaries (grey bars) and an unbiased comparison binary
sample \citep{duquennoy1991} with G type primaries (white bars).}\label{periods}
\end{figure}

With the masses and semi-major axes from Table\,\ref{table8}, we can derive the orbital periods of
all planet host stellar systems and compare their distribution with the period distribution of an
unbiased comparison binary sample with G type primaries \citep{duquennoy1991}, which are both shown
in Fig.\,\ref{periods}. We determined the semi-major axes of Class-1 systems with the relation
between semi-major axes $a$ and projected separations $sep$ also used by \citet{duquennoy1991}
($log(a/sep)=0.13$). For Class-2 systems, we used the given orbital periods derived from orbit
fitting. Apparently, there is a lack of close planet host stellar systems. Most of the detected
planet host binaries are found with orbital periods $6<log(P[days])<7$ while the unbiased control
sample exhibits its peak in the period bin $4<log(P[days])<5$.

The difference between the two period distributions is probably the result of an observational
selection effect, due to the difficulty detecting planets in close binary systems using the
radial-velocity technique. If we assume a limit-angular separation of 2.5\,arcsec for the planet
search in binaries, this yields a limiting orbital period of $log(P[days]\sim5.6$, calculated with
the average distance of the exoplanet host stars (40\,pc) and a binary total mass of 1\,$M_{\sun}$.
Indeed, we count only 6 planet host binaries with shorter orbital periods and 23 with longer
orbital periods. Most of these close systems are composed of a bright exoplanet host star and a
much fainter secondary star that does not disturb the radial-velocity planet search technique. Only
a systematic planet-search in wide and close binary systems would be able to indicate whether
planet host binaries tend to be more widely separated than binaries without planets.

\begin{figure} [h]
\resizebox{\hsize}{!}{\includegraphics{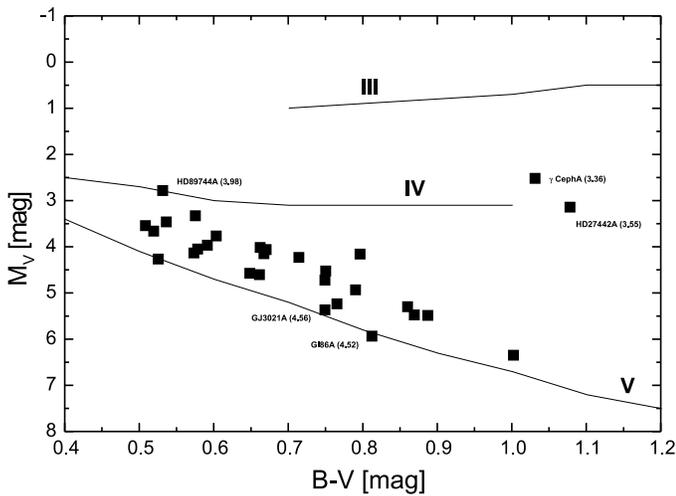}}\caption{Hipparcos photometry of all
exoplanet host stars located in multiple-star systems. The absolute V-band magnitudes are obtained
with distances derived from the Hipparcos parallaxes. For comparison we show the expected colors
and magnitudes of dwarfs, subgiants, and giants taken from \citet{schmidtkaler1982}.}\label{fig17}
\end{figure}

Most of the exoplanet host stars that reside in multiple-star systems are dwarfs with spectral
types between late F and early K. Their properties are well-known, derived from spectroscopy and
astrometry. The location of these stars in a color-magnitude diagram is shown in Fig.\,\ref{fig17}.
The absolute V-band magnitudes are determined with distances from Hipparcos \citep{perryman1997},
which also provides the V- and B-band photometry. For comparison, we plot continuous lines that
present the intrinsic colors and magnitudes of dwarfs, subgiants, and giants \citep[data
from][]{schmidtkaler1982}. The brightest and most evolved stars in the sample are HD\,89744\,A,
$\gamma$\,Cep\,A, and HD\,27442\,A.

HD\,89744\,A is classified as an F7V dwarf. However, both its low surface-gravity
log(g[$cm/s^{2}$])\,=\,3.98$\pm$0.05 \citep{santos2004} and its position in the color-magnitude
diagram are only marginally consistent with the dwarf classification, which indicates that this
star has already evolved.

The planet host star $\gamma$\,Cep\,A \citep{hatzes2003} is listed in the Hipparcos catalogue as a
subgiant with a spectral type K1IV. \citet{santos2004} derive a surface gravity of
log(g[$cm/s^{2}$])\,=\,3.36$\pm$0.21, which is consistent with the subgiant classification. The
star $\gamma$\,Cep is a single-lined spectroscopic binary system \citep{campbell1988}, whose
secondary, $\gamma$\,Cep\,B, was also recently detected directly by \cite{neuhaeuser2007}.

As described in Sect.\,2.2, HD\,27442\,A is an evolved early K type subgiant. The derived spectral
type and its luminosity class are both fully consistent with its position in the color-magnitude
diagram in Fig.\,\ref{fig17}. Furthermore, this binary system is a special case among all known
planet host
 binary systems. With its subgiant exoplanet host star and the white dwarf companion
HD\,27442\,B, it is the most evolved planet host stellar system known today.

\section{Discussion}

\subsection{Long-term stability of exoplanets in multiple-star systems}

According to \citet{holman1999} planets in binary systems are stable on a long timescale only if
their semi-major axes (circular orbit assumed) do not extend a critical value, the so-called
critical semi-major axis $a_{c}$. This axis depends on the mass-ratio of the primary and secondary
($\mu$\,=\,$m_{Sec}/(m_{Prim}+m_{Sec})$), as well as on the binary eccentricity. We followed
\citet{holman1999} prescription and derived in Table\,\ref{table8} the critical semi-major axis for
all the multiple-star systems with exoplanets.

The derived critical semi-major axes range from 1\,AU (HD\,188753) up to 800\,AU (HD\,40979). They
can be used to estimate the long-term stable regions of additional planets or brown-dwarf
companions in these systems around the planet host stars. Only four planet host multiples are
presently known to have critical semi-major axes below 5\,AU. These are HD\,188753 (1\,AU),
HD\,41004 (3\,AU), Gl\,86 (3\,AU), and $\gamma$\,Cep (4\,AU). Note that all four exoplanets reside
within the proposed long-term stable regions. However, the existence of gaseous giant, Jovian-like
planets on orbits similar to planets in our own solar system (a\,$>$\,5\,AU) are not stable in
these systems. Furthermore, Uranus- or Neptune-like orbits, with a separation between 20 and
30\,AU, as well as Kuiper belts that extend beyond 20\,AU, can be ruled out in about a third of all
planet host stellar systems.

The extent of the long-term stable regions changes during the post-main sequence evolution of the
stellar system. Due to the mass-loss of the stars, the binary mass-ratio $\mu$ and its semi-major
axis change lead to a variation of the long-term stable regions around the individual stars. An
interesting example of this effect is the binary system HD\,27442\,AB, a wider analog to the
Gl\,86\,AB system \citep[see Mugrauer et al. 2005a\nocite{mugrauer2005a}; ][]{lagrange2006}. With a
white dwarf companion and a subgiant planet host star, it is the most evolved planet host binary
system known today. According to the initial to final mass-relation from \citet{weidemann2000}, the
white-dwarf progenitor was most probably a solar-like star. Due to the mass-loss during the
red-giant phase of the white-dwarf progenitor, the binary semi-major axis had to expand slightly by
a factor of $\sim$1.7, i.e. from $\sim$140\,AU to the current value of $\sim$240\,AU. Due to these
variations the long-term stable region around the exoplanet host star extended from 14\,AU up to
32\,AU. The known exoplanet that revolves around HD\,27442\,A in its 1.2\,AU circular orbit was
well within the smaller long-term stable region and was therefore not dynamically affected, even
before the mass-loss of the secondary star.

The known exoplanet survived the mass-loss of HD\,27442\,B but might not survive the evolution of
its parent star. HD\,27442\,A is already slightly evolved and will pass through the final stages of
stellar evolution in the next few hundred million years. \citet{livio1984} showed that, during the
red-giant phase of its parent star, a close enough planet spirals inwards, which results in
evaporation of the planet atmosphere that eventually dissolves the planet in only a few thousand
years. Therefore, the presently known planet probably will not last for more than a few millon
years. Other planets in the system with separations larger than $\sim$5\,AU might survive the
red-giant stage \citep{burleigh2002} while their orbits expand by a factor of two. Their orbit will
still remain within the long-term stable region around HD\,27442\,A. Also, HD\,27442\,B might have
wide survivor planets, orbiting the white dwarf in its long-term stable of about 20\,AU
(1\,arcsec).

\subsection{Statistical differences between single-star and multiple-star planets}

Differences in the mass-period and eccentricity-period correlation between the single- and
multiple-star planets have been already pointed out by \citet{zucker2002b} and
\citet{eggenberger2004}. These authors note that in the mass-period diagram for orbital periods
shorter than 40 days, massive planets ($msin(i)$\,$>$\,2\,$M_{Jup}$) are only found among the
multiple-star planets, whereas all the detected single-star planets have lower masses. In contrast,
multiple-star planets exhibit only masses lower than 5\,$M_{Jup}$ for orbital periods longer than
100 days, whereas the masses of single-star planets clearly extend beyond this mass-limit. The two
populations seem different in the eccentricity-period diagram as well. For orbital periods shorter
than 40 days multiple-star planets revolve around their parent stars only on almost circular orbits
(e\,$\leq$\,0.05), whereas single-star planets could have higher orbital eccentricities, up to
$\sim$0.5. We can check these conjectures now with the updated enlarged dataset.

As listed in Table\,\ref{table8}, 35 of the exoplanets are known to orbit a star that is a member
of binary or even triple-star systems. Among those stellar systems, we also find apparently wide
systems with approximated semi-major axes of more than 2000\,AU (GJ\,777, HD\,2132140, HD\,222582,
HD\,89744, and HD\,40979). From the theoretical point of view, it might be difficult to imagine the
influence of those wide companions on the formation and orbital evolution of exoplanets, even with
the Kozai effect and migration scenario. However, as already described in the previous section, the
orbital elements are only approximated (Class-1 systems) for the majority of these stellar systems.
Therefore, it might be possible that these apparently wide systems exhibit much smaller semi-major
axes with high orbital eccentricities. Furthermore, presently wide binary systems might have been
much closer in the past, because the orbital elements of the binary systems could also have been
altered, e.g., due to close encounters with other stars during the system formation either in a
stellar cluster or later in the galactic plane. Note also that evolved systems with white-dwarf
companions have expanded their semi-major axes due to mass-loss of the white-dwarf progenitors. We
therefore decided to analyze all multiple-star planets as one group and compare them with the
single-star planets.

To avoid selection effects we discarded all planets with small radial-velocity amplitudes
$K$\,$<$\,15\,m/s. In particular 5 multiple-star planets, namely 55\,Cnc\,e and c, 83\,Leo\,b,
GJ\,777\,c, and HD\,16141\,b, are not considered in the analysis because of their low
radial-velocity amplitudes. Altogether we had 108 single-star and 30 multiple-star planets by the
end of 2005.

\begin{figure} [t]
\includegraphics[width=8cm]{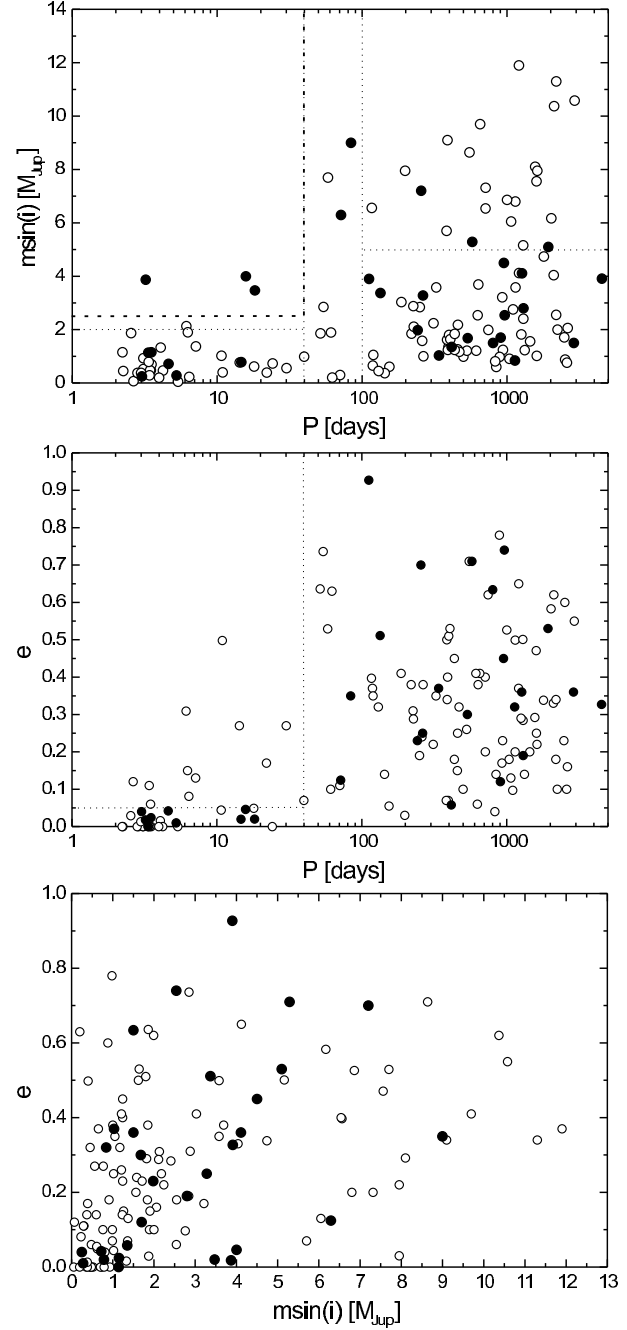}\caption{The mass-period (top), eccentricity-period (middle),
and eccentricity-mass (bottom) correlation diagrams of single-star (open circles) and multiple-star
planets (filled circles). The mass and eccentricity-limits, as proposed by \citet{eggenberger2004},
are illustrated with dotted lines. The updated mass-limit for short period planets is shown as a
thick dashed line. Only planets with $msin(i)$\,$<$\,13\,$M_{Jup}$ and $K$\,$>$\,15\,m/s are taken
into account.}\label{fig15}
\end{figure}

The correlation diagrams of both planet populations are shown in Fig.\,\ref{fig15}, together with
the mass- and eccentricity-limits for short and long-period planets as proposed by
\citet{eggenberger2004} (see dotted lines in Fig.\,\ref{fig15}). In the updated planet sample, the
mass-limit has to be increased from 2 to 2.5\,$M_{Jup}$ due to the single-star planet
HD\,188203\,b, which slightly extends the mass-limit proposed by \cite{eggenberger2004}.

In the short-period domain, P\,$<$\,40 days, we count 30 single-star and 6 multiple-star planets
with $msin(i)$\,$\leq$\,2.5\,$M_{Jup}$. There are 3 multiple-star planets with
$msin(i)$\,$>$\,2.5\,$M_{Jup}$ and no single-star planet. If we assume that both planet populations
have the same mass-period distribution, the probability of the given planet distribution can be
derived with a hypergeometric distribution, which yields a probability of
0.92\,\%\footnote{Hypergeometric distribution:
$\left(\begin{array}{c}36\\6\\\end{array}\right)\left(\begin{array}{c}3\\3\\\end{array}\right)\,/\,\left(\begin{array}{c}39\\9\\\end{array}\right)$\,=\,0.92\,\%\newline}.

\citet{eggenberger2004} also suggest that the planets in multiple-star systems with orbital periods
longer than 100 days are less massive. However, in the updated plot, three multiple-star planets
(HD\,89744\,b, HD\,222582\,b, HD\,142022\,b) exceed the proposed mass-limit
($msin(i)$\,$<$\,5\,$M_{Jup}$) for the given range of orbital periods. These outliers substantially
weaken the significance of the suggested feature.

\begin{figure} [t]
\includegraphics[width=8cm]{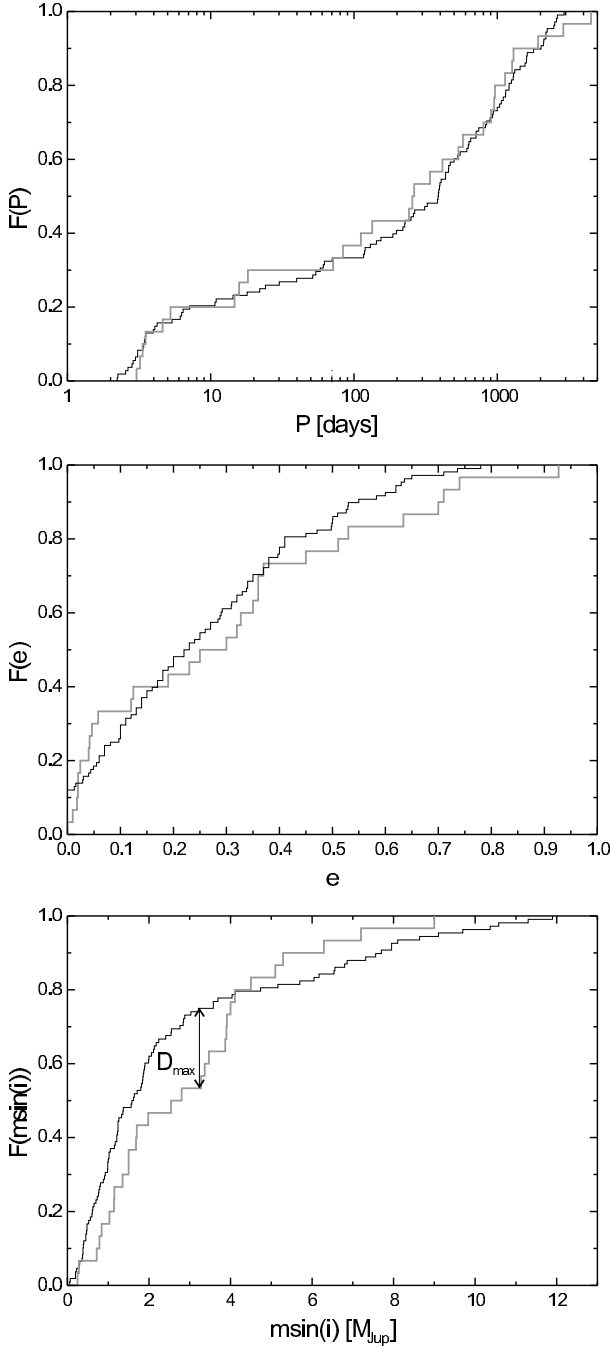}\caption{The cumulative functions of planet
periods (top), eccentricities (middle), and masses (bottom). The functions of single-star planets
are shown as thin black lines and those of multiple-star planets as thick grey lines. All functions
are scaled equally on the probability axis; i.e. the three functions can be directly compared with
each other. The largest detected difference D$_{max}$ between both planet populations occurs in the
cumulative mass-functions and is indicated by a black line with arrows. Only planets with
$msini(i)$\,$<$\,13\,$M_{Jup}$ and $K$\,$>$\,15\,m/s are taken into account.}\label{fig16}
\end{figure}

In the updated eccentricity-period diagram, we count 18 single-star planets and 9 multiple-star
planets all on almost circular orbits (e\,$\leq$\,0.05), and 12 single-star planets with
e\,$>$\,0.05 all with periods shorter than 40 days. There is no multiple-star planet found on an
eccentric orbit for the given range of orbital periods. If we assume that there is no difference in
the eccentricity-period distribution of both exoplanet populations, the probability for the given
planet distribution is 2.21\,\%\footnote{Hypergeometric distribution:
$\left(\begin{array}{c}27\\9\\\end{array}\right)\left(\begin{array}{c}12\\0\\\end{array}\right)\,/\,\left(\begin{array}{c}39\\9\\\end{array}\right)$\,=\,2.21\,\%\newline}.
This result is similar to the 3.77\,\%, reported by \citet{eggenberger2004}. The remaining
correlation of the planet properties, namely the eccentricity-mass correlation is shown in the
bottom panel of Fig.\,\ref{fig15}. The mass-eccentricity distributions of both planet populations
seem similar, with no significant statistical difference.

In addition to the analysis of the correlation diagrams, we also compared the period, eccentricity,
and mass-distribution functions of both planet populations. The three cumulative distribution
functions $F$ are plotted in Fig.\,\ref{fig16}. The cumulative functions of the orbital parameters
(period $F(P)$ and eccentricity $F(e)$) of both planet populations appear to be more or less
identical with only small variations. In contrast, the mass-distributions $F(msin(i))$ seem
different, although Kolmogorov-Smirnov (K-S) two-sample test indicates that the hypothesis that the
two distributions are equal can be rejected with only 81\% significance level. The biggest
discrepancy ($D_{max}$) between both cumulative mass-functions appears at about 3\,$M_{Jup}$ (see
Fig.\,\ref{fig16}).

In a further step in examining the possible difference between the mass-distributions, we derived
the underlying density functions from the cumulative distribution functions. To derive these
function we interpolated the cumulative mass-distribution functions of both planet populations
using 1000 data points spanning a range of mass from 0 to 13\,$M_{Jup}$. We apply an adjacent
averaging with a boxwidth of 50 data points. The value $y_{i}$ for each point $x_{i}$ is the
average of the values of all points in the interval $[x_{i-25};x_{i+25}]$.

\begin{figure} [h]
\includegraphics[width=8cm]{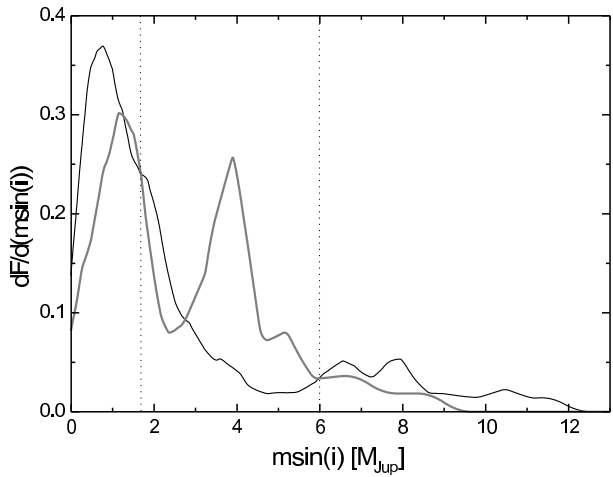}\caption{The derived mass-distribution of single-star
planets (thin black line) and multiple-star planets (thick grey line). The mass-distribution
functions are derived from the cumulative mass-distribution functions (see text for more details).
More massive single-star planets are less probable, whereas multiple-star planets exhibit another
peak in their mass-distribution at about 4\,$M_{Jup}$.}\label{fig16b}
\end{figure}

The derivative is taken by averaging the two adjacent data slopes at point $x_{i}$:
$f'(x_{i})=\frac{1}{2}\left(\frac{y_{i+1}-y_{i}}{x_{i+1}-x_{i}}+\frac{y_{i}-y_{i-1}}{x_{i}-x_{i-1}}\right)$.
High spatial frequencies are smoothed again with an adjacent average with a boxwidth of 50 data
points. The derived mass-distribution functions are shown in Fig.\,\ref{fig16b}.

Both planet populations exhibit a similar mass-distribution in the low-mass and high-mass range,
with a high peak at about 1\,$M_{Jup}$, but they differ at the intermediate mass-range, between
$\sim$2 and 6\,$M_{Jup}$. At this range, the multiple-star planets seem to be more frequent than
single-star planets. If this difference is true, it might indicate that the planet formation
process is indeed altered by the stellar multiplicity. One possibility is that the mass-accretion
rate onto the forming planet embedded in a protoplanetary disk is enhanced \citep[as described
by][]{kley2000}. Another possibility is that, because of the angular momentum transfer from the
planet bearing disks to the perturbing companion star, more material resides closer in, which
finally allows the formation of more massive planets.

Finally, it is important to mention that all the reported possible differences between the
properties of single- and multiple-star planets have to be reexamined when the sample size of
planet host stellar systems is significantly extended, at least by a factor of two, i.e. when about
60 such systems are known. While writing the manuscript of this paper, new companions or
interesting companion candidates already had been presented (see e.g. Chauvin et. al.
2006\nocite{chauvin2006}, Raghavan et al. 2006\nocite{raghavan2006} and Bakos et al.
2006\nocite{bakos2006}). In fact, we also plan to show new promising results of our multiplicity
study in a forthcoming publication.

\section{Summary}

\begin{itemize}
\item[1.] GJ\,3021\,B is a close M4V-M5V stellar companion, as confirmed by ISAAC
spectroscopy\newline
\item[2.] The wide companion of the binary system HD\,40979\,AB is resolved in a pair B+C; i.e. this is a new
triple-star system that host a planet\newline
\item[3.] HD\,27442\,B is a co-moving companion of an exoplanet host star
whose white dwarf nature, is confirmed by ISAAC and FORS spectroscopy.\newline
\item[4.] HD\,27442\,AB is the most evolved planet host stellar system known today.\newline
\item[5.] Massive (msin(i)\,$>$\,2.5\,$M_{Jup}$) shortperiod planets (P\,$<$\,40\,days) apparently all
reside in multiple-star systems, probably on almost circular orbits (e\,$\leq$\,0.05).\newline
\item[6.] Intermediate-mass planets ($\sim$4$\pm$2\,$M_{Jup}$) might be more frequent in multiple-star systems
than around single stars.
\end{itemize}

\acknowledgements {We would like to thank the technical staff of the ESO NTT and Joint Astronomy
Center, Hawaii, for help and assistance in carrying out the observations. We thank A.Seifahrt and
A.~Szameit who carried out some of the observations at UKIRT, and G.~F\"orster for his assistance
during the observations, at Wendelstein observatory. We thank Eike Guenther of the Th\"uringer
Landessternwarte Tautenburg for all his help. Furthermore, we thank H.~Barwig, O.~B\"arnbantner,
J.~Koppenh\"ofer, and W.~Mitsch of the Munich LMU university observatory for all their support.
Finally, the authors want to thank Ga\"el Chauvin (A\&A referee) for his comments and suggestions.
We made use of the 2MASS public data releases, as well as the Simbad database operated at the
Observatoire Strasbourg. T.M. thanks the Israel Science Foundation for support through grant no.
03/233.}

{}


\begin{thebibliography}{}

\bibitem[Allende Prieto \& Lambert(1999)]{allende1999} Allende Prieto, C., \& Lambert, D.~L.\ 1999, \aap, 352, 555
\bibitem[Bakos et al.(2006)]{bakos2006} Bakos, G.~{\'A}., P{\'a}l, A., Latham, D.~W., Noyes, R.~W., \& Stefanik, R.~P.\ 2006, \apjl, 641, L57
\bibitem[Baraffe et al.(1998)]{baraffe1998} Baraffe, I., Chabrier, G., Allard, F., \& Hauschildt, P.~H.\ 1998, \aap, 337, 403
\bibitem[Baraffe et al.(2003)]{baraffe2003} Baraffe, I., Chabrier, G., Barman, T.~S., Allard, F., \& Hauschildt, P.~H.\ 2003, \aap, 402, 701
\bibitem[Bergeron et al.(2001)]{bergeron2001} Bergeron, P., Leggett, S.~K., \& Ruiz, M.~T.\ 2001, \apjs, 133, 413
\bibitem[Burleigh et al.(2002)]{burleigh2002} Burleigh, M.~R., Clarke, F.~J., \& Hodgkin, S.~T.\ 2002, \mnras, 331, L41
\bibitem[Butler et al.(1997)]{butler1997} Butler, R.~P., Marcy, G.~W., Williams, E., Hauser, H., \& Shirts, P.\ 1997, \apjl, 474, L115
\bibitem[Butler et al.(1999)]{butler1999} Butler, R.~P., Marcy, G.~W., Fischer, D.~A., Brown, T.~M., Contos, A.~R., Korzennik, S.~G., Nisenson, P., \& Noyes, R.~W.\ 1999, \apj, 526, 916
\bibitem[Butler et al.(2001)]{butler2001} Butler, R.~P., Tinney, C.~G., Marcy, G.~W., Jones, H.~R.~A., Penny, A.~J., \& Apps, K.\ 2001, \apj, 555, 410
\bibitem[Butler et al.(2003)]{butler2003} Butler, R.~P., Marcy, G.~W., Vogt, S.~S., Fischer, D.~A., Henry, G.~W., Laughlin, G., \& Wright, J.~T.\ 2003, \apj, 582, 455
\bibitem[Campbell et al.(1988)]{campbell1988} Campbell, B., Walker, G.~A.~H., \& Yang, S.\ 1988, \apj, 331, 902
\bibitem[Chauvin et al.(2006)]{chauvin2006} Chauvin, G., Lagrange, A.-M., Udry, S., Fusco, T., Galland, F., Naef, D., Beuzit, J.-L., \& Mayor, M.\ 2006, \aap, 456, 1165
\bibitem[Cochran et al.(1997)]{cochran1997} Cochran, W.~D., Hatzes, A.~P., Butler, R.~P., \& Marcy, G.~W.\ 1997, \apj, 483, 457
\bibitem[Cushing et al.(2005)]{cushing2005} Cushing, M.~C., Rayner, J.~T., \& Vacca, W.~D.\ 2005, \apj, 623, 1115
\bibitem[Cutri et al.(2003)]{cutri2003} Cutri, R.~M., et al.\ 2003, The IRSA 2MASS All-Sky Point Source Catalog, NASA/IPAC Infrared Science Archive.~http://irsa.ipac.caltech.edu/applications/Gator/,
\bibitem[Danks \& Dennefeld(1994)]{danks1994} Danks, A.~C., \& Dennefeld, M.\ 1994, \pasp, 106, 382
\bibitem[Desidera et al.(2004)]{desidera2004} Desidera, S., et al.\ 2004, ASP Conf.~Ser.~321: Extrasolar Planets: Today and Tomorrow, 321, 103
\bibitem[Devillard(2001)]{devillard2001} Devillard, N.\ 2001, ASP Conf.~Ser.~238: Astronomical Data Analysis Software and Systems X, 238, 525
\bibitem[Duquennoy \& Mayor(1991)]{duquennoy1991} Duquennoy, A., \& Mayor, M.\ 1991, \aap, 248, 485
\bibitem[Eggenberger et al.(2004)]{eggenberger2004} Eggenberger, A., Udry, S., \& Mayor, M.\ 2004, \aap, 417, 353
\bibitem[Eggenberger et al.(2006)]{eggenberger2006} Eggenberger, A., Mayor, M., Naef, D., Pepe, F., Queloz, D., Santos, N.~C., Udry, S., \& Lovis, C.\ 2006, \aap, 447, 1159
\bibitem[Els et al.(2001)]{els2001} Els, S.~G., Sterzik, M.~F., Marchis, F., Pantin, E., Endl, M., K\"urster, M.\ 2001, \aap, 370, L1
\bibitem[Fischer et al.(2003)]{fischer2003} Fischer, D.~A., et al.\ 2003, \apj, 586, 1394
\bibitem[Fischer et al.(2006)]{fischer2006} Fischer, D.~A., et al.\ 2006, \apj, 637, 1094
\bibitem[Ford et al.(2000)]{ford2000} Ford, E.~B., Kozinsky, B., \& Rasio, F.~A.\ 2000, \apj, 535, 385
\bibitem[Fuhrmann(2004)]{fuhrmann2004} Fuhrmann, K.\ 2004, Astronomische Nachrichten, 325, 3
\bibitem[Halbwachs(1986)]{halbwachs1986} Halbwachs, J.~L.\ 1986, \aaps, 66, 131
\bibitem[Hartkopf(2005)]{hartkopfonline} Hartkopf, W.~I., Mason, B.~M. and Worley, C.~E.\ 2005 , Sixth Catalog of Orbits of Visual Binary Stars, http://ad.usno.navy.mil/wds/orb6/orb6.html
\bibitem[Hatzes et al.(2003)]{hatzes2003} Hatzes, A.~P., Cochran, W.~D., Endl, M., McArthur, B., Paulson, D.~B., Walker, G.~A.~H., Campbell, B., \& Yang, S.\ 2003, \apj, 599, 1383
\bibitem[Holman \& Wiegert(1999)]{holman1999} Holman, M.~J., \& Wiegert, P.~A.\ 1999, \aj, 117, 621
\bibitem[Jahrei{\ss}(2001)]{jahreiss2001} Jahrei{\ss}, H.\ 2001, Astronomische Gesellschaft Meeting Abstracts, 18, 110
\bibitem[Jones et al.(2002)]{jones2002} Jones, H.~R.~A., Paul Butler, R., Marcy, G.~W., Tinney, C.~G., Penny, A.~J., McCarthy, C., \& Carter, B.~D.\ 2002, \mnras, 337, 1170
\bibitem[Kenyon \& Hartmann(1995)]{kenyon1995} Kenyon, S.~J., \& Hartmann, L.\ 1995, \apjs, 101, 117
\bibitem[Kley(2000)]{kley2000} Kley, W.\ 2000, IAU Symposium, 200, 211P
\bibitem[Konacki(2005a)]{konacki2005a} Konacki, M.\ 2005a, \nat, 436, 230
\bibitem[Konacki(2005b)]{konacki2005b} Konacki, M.\ 2005b, \apj, 626, 431
\bibitem[Korzennik et al.(2000)]{korzennik2000} Korzennik, S.~G., Brown, T.~M., Fischer, D.~A., Nisenson, P., \& Noyes, R.~W.\ 2000, \apjl, 533, L147
\bibitem[Lagrange et al.(2006)]{lagrange2006} Lagrange, A.-M., Beust, H., Udry, S., Chauvin, G., \& Mayor, M.\ 2006, \aap, 459, 955
\bibitem[Liebert et al.(2005)]{liebert2005} Liebert, J., Bergeron, P., \& Holberg, J.~B.\ 2005, \apjs, 156, 47
\bibitem[Livio \& Soker(1984)]{livio1984} Livio, M., \& Soker, N.\ 1984, \mnras, 208, 763
\bibitem[Lowrance et al.(2002)]{lowrance2002} Lowrance, P.~J., Kirkpatrick, J.~D., \& Beichman, C.~A.\ 2002, \apjl, 572, L79
\bibitem[Marcy et al.(2000)]{marcy2000} Marcy, G.~W., Butler, R.~P., \& Vogt, S.~S.\ 2000, \apjl, 536, L43
\bibitem[Marcy et al.(2005)]{marcy2005} Marcy, G.~W., Butler, R.~P., Vogt, S.~S., Fischer, D.~A., Henry, G.~W., Laughlin, G., Wright, J.~T., \& Johnson, J.~A.\ 2005, \apj, 619, 570
\bibitem[Marzari \& Scholl(2000)]{marzari2000} Marzari, F., \& Scholl, H.\ 2000, \apj, 543, 328
\bibitem[Mayer et al.(2005)]{mayer2005} Mayer, L., Wadsley, J., Quinn, T., \& Stadel, J.\ 2005, \mnras, 363, 641
\bibitem[Mayor \& Queloz(1995)]{mayor1995} Mayor, M., \& Queloz, D.\ 1995, \nat, 378, 355
\bibitem[Mayor et al.(2004)]{mayor2004} Mayor, M., Udry, S., Naef, D., Pepe, F., Queloz, D., Santos, N.~C., \& Burnet, M.\ 2004, \aap, 415, 391
\bibitem[Mazeh et al.(1996)]{mazeh1996} Mazeh, T., Latham, D.~W., \& Stefanik, R.~P.\ 1996, \apj, 466, 415
\bibitem[McArthur et al.(2004)]{mcarthur2004} McArthur, B.~E., et al.\ 2004, \apjl, 614, L81
\bibitem[Mugrauer et al.(2004)]{mugrauer2004a} Mugrauer, M., Neuh{\"a}user, R., Mazeh, T., Alves, J., \& Guenther, E.\ 2004a, \aap, 425, 249
\bibitem[Mugrauer et al.(2004)]{mugrauer2004b} Mugrauer, M., Neuh{\"a}user, R., Mazeh, T., Guenther, E., \& Fern{\'a}ndez, M.\ 2004b, Astronomische Nachrichten, 325, 718
\bibitem[Mugrauer \& Neuh{\"a}user(2005)]{mugrauer2005a} Mugrauer, M., \& Neuh{\"a}user, R.\ 2005, \mnras, 361, L15
\bibitem[Mugrauer et al.(2005)]{mugrauer2005b} Mugrauer, M., Neuh{\"a}user, R., Seifahrt, A., Mazeh, T., \& Guenther, E.\ 2005, \aap, 440, 1051
\bibitem[Mugrauer et al.(2006)]{mugrauer2006} Mugrauer, M., Neuh{\"a}user, R., Mazeh, T., Guenther, E., Fern{\'a}ndez, M., \& Broeg, C.\ 2006, Astronomische Nachrichten, 327, 321
\bibitem[Naef et al.(2001)]{naef2001a} Naef, D., et al.\ 2001a, \aap, 375, L27
\bibitem[Naef et al.(2001)]{naef2001b} Naef, D., Mayor, M., Pepe, F., Queloz, D., Santos, N.~C., Udry, S., \& Burnet, M.\ 2001b, \aap, 375, 205
\bibitem[Neuh{\"a}user et al.(2007)]{neuhaeuser2007} Neuh{\"a}user, R., Mugrauer, M., Fukagawa, M., Torres, G., \& Schmidt, T.\ 2007, \aap, 462, 777
\bibitem[Patience et al.(2002)]{patience2002} Patience, J., et al.\ 2002, \apj, 581, 654
\bibitem[Perryman \& ESA(1997)]{perryman1997} Perryman, M.~A.~C., \& ESA 1997, The Hipparcos and Tycho catalogues.~Astrometric and photometric star catalogues derived from the ESA Hipparcos Space Astrometry Mission, Publisher: Noordwijk, Netherlands: ESA Publications Division, 1997, Series: ESA SP Series vol no: 1200, ISBN: 9290923997 (set),
\bibitem[Pichardo et al.(2005)]{pichardo2005} Pichardo, B., Sparke, L.~S., \& Aguilar, L.~A.\ 2005, \mnras, 359, 521
\bibitem[Pickles(1998)]{pickles1998} Pickles, A.~J.\ 1998, \pasp, 110, 863
\bibitem[Queloz et al.(2000)]{queloz2000} Queloz, D., et al.\ 2000, \aap, 354, 99
\bibitem[Raghavan et al.(2006)]{raghavan2006} Raghavan, D., Henry, T.~J., Mason, B.~D., Subasavage, J.~P., Jao, W.-C., Beaulieu, T.~D., \& Hambly, N.~C.\ 2006, \apj, 646, 523
\bibitem[Randich et al.(1999)]{randich1999} Randich, S., Gratton, R., Pallavicini, R., Pasquini, L., \& Carretta, E.\ 1999, \aap, 348, 487
\bibitem[Rocha-Pinto \& Maciel(1998)]{rocha1998} Rocha-Pinto, H.~J., \& Maciel, W.~J.\ 1998, \mnras, 298, 332
\bibitem[Saffe et al.(2005)]{saffe2005} Saffe, C., G{\'o}mez, M., \& Chavero, C.\ 2005, \aap, 443, 609
\bibitem[Santos et al.(2001)]{santos2001} Santos, N.~C., Mayor, M., Naef, D., Pepe, F., Queloz, D., Udry, S., \& Burnet, M.\ 2001, \aap, 379, 999
\bibitem[Santos et al.(2004)]{santos2004} Santos, N.~C., Israelian, G., \& Mayor, M.\ 2004, \aap, 415, 1153
\bibitem[Schmidt-Kaler(1982)]{schmidtkaler1982} Schmidt-Kaler, T.\ 1982, Landolt-B\"ornstein, ed. K. Schaifers and H.H. Voigt, Vol.2 (Berlin: Springer), 499
\bibitem[S{\"o}derhjelm(1999)]{soederhjelm1999} S{\"o}derhjelm, S.\ 1999, \aap, 341, 121
\bibitem[Tinney et al.(2002)]{tinney2002} Tinney, C.~G., Butler, R.~P., Marcy, G.~W., Jones, H.~R.~A., Penny, A.~J., McCarthy, C., \& Carter, B.~D.\ 2002, \apj, 571, 528
\bibitem[Tokovinin et al.(2000)]{tokovinin2000} Tokovinin, A.~A., Griffin, R.~F., Balega, Y.~Y., Pluzhnik, E.~A., \& Udry, S.\ 2000, Astronomy Letters, 26, 116
\bibitem[Toyota et al.(2005)]{toyota2005} Toyota, E., Itoh, Y., Matsuyama, H., Urakawa, S., Kimura, S., Oasa, Y., Mukai, T., \& Sato, B.\ 2005, Protostars and Planets V, 8247
\bibitem[Udry et al.(2000)]{udry2000} Udry, S., et al.\ 2000, \aap, 356, 590
\bibitem[Vogt et al.(2000)]{vogt2000} Vogt, S.~S., Marcy, G.~W., Butler, R.~P., \& Apps, K.\ 2000, \apj, 536, 902
\bibitem[Vogt et al.(2005)]{vogt2005} Vogt, S.~S., Butler, R.~P., Marcy, G.~W., Fischer, D.~A., Henry, G.~W., Laughlin, G., Wright, J.~T., \& Johnson, J.~A.\ 2005, \apj, 632, 638
\bibitem[Wallace \& Hinkle(1997)]{wallace1997} Wallace, L., \& Hinkle, K.\ 1997, \apjs, 111, 445
\bibitem[Weidemann(2000)]{weidemann2000} Weidemann, V.\ 2000, \aap, 363, 647
\bibitem[Worley \& Douglass(1997)]{worley1997} Worley, C.~E., \& Douglass, G.~G.\ 1997, \aaps, 125, 523
\bibitem[Wu \& Murray(2003)]{wu2003} Wu, Y., \& Murray, N.\ 2003, \apj, 589, 605
\bibitem[Zacharias et al.(2004)]{zacharias2004} Zacharias, N., Urban, S.~E., Zacharias, M.~I., Wycoff, G.~L., Hall, D.~M., Monet, D.~G., \& Rafferty, T.~J.\ 2004, \aj, 127, 3043
\bibitem[Zucker et al.(2002)]{zucker2002a} Zucker, S., et al.\ 2002, \apj, 568, 363
\bibitem[Zucker \& Mazeh(2002)]{zucker2002b} Zucker, S., \& Mazeh, T.\ 2002, \apjl, 568, L113
\bibitem[Zucker et al.(2003)]{zucker2003} Zucker, S., Mazeh, T., Santos, N.~C., Udry, S., \& Mayor, M.\ 2003, \aap, 404, 775
\bibitem[Zucker et al.(2004)]{zucker2004} Zucker, S., Mazeh, T., Santos, N.~C., Udry, S., \& Mayor, M.\ 2004, \aap, 426, 695





























































\end{thebibliography}
\end{document}